\documentclass{article}

\usepackage[utf8]{inputenc}
\usepackage[hyphens]{url}
\usepackage{lipsum}
\usepackage{graphicx}
\usepackage{amssymb,amsthm,amsmath}
\mathchardef\mhyphen="2D
\usepackage{epstopdf}
\usepackage{pdfsync}
\usepackage[margin=1.5in]{geometry}

\usepackage[round,numbers]{natbib}

\usepackage{todonotes}

\usepackage[utf8]{inputenc}
\usepackage[T1]{fontenc}
\usepackage{newtxtext,newtxmath}

\usepackage{braket}
\usepackage{physics}
\usepackage{caption}
\usepackage{pdfpages}

\newcommand{\beq}{\begin{equation}}
\newcommand{\eeq}{\end{equation}}
\newcommand{\beqn}{\begin{equation*}}
\newcommand{\eeqn}{\end{equation*}}
\newcommand{\of}[1]{\left(#1\right)}

\newcommand{\kin}{\kappa_{in}}

\newcommand{\ybyvo}{$^{171}$Yb:YVO~}

\newcommand{\yvo}{YVO}

\newcommand{\ybion}{ ${}^{171}\mathrm{Yb}^{3+}$~}

\newcommand{\eq}[1]{Eq.~(\ref{eq:#1})}
\newcommand{\fig}[1]{Fig.~\ref{fig:#1}}

\newcommand{\sect}[1]{Section~\ref{sec:#1}}

\title{Supplementary materials: Coherent control and single-shot readout of a rare-earth ion embedded in a nanophotonic cavity}

\date{}

\begin{document}


# Coherent control and single-shot readout of a rare-earth ion embedded in a nanophotonic cavity

Jonathan M. Kindem[1,2], Andrei Ruskuc[1,2], John G. Bartholomew[1,2], Jake Rochman[1,2], Yan Qi Huan[1,2], Andrei Faraon[1,2,*]

**Affiliations:**

[1]Kavli Nanoscience Institute and Thomas J. Watson, Sr., Laboratory of Applied Physics, California Institute of Technology, Pasadena, California 91125, USA.

[2]Institute for Quantum Information and Matter, California Institute of Technology, Pasadena, California 91125, USA.

*Correspondence to: faraon@caltech.edu

**Abstract:**
Quantum networks based on optically addressable spin qubits promise to enable secure communication, distributed quantum computing, and tests of fundamental physics. Scaling up quantum networks based on solid-state luminescent centers requires coherent spin and optical transitions coupled to photonic resonators. Here we investigate single $^{171}$Yb$^{3+}$ ions in yttrium orthovanadate coupled to a nanophotonic cavity. These ions possess optical and spin transitions that are first-order insensitive to magnetic field fluctuations, enabling optical linewidths less than 1 MHz and spin coherence times exceeding 30 ms for cavity-coupled ions. The cavity-enhanced optical emission rate facilitates efficient spin initialization and conditional single-shot readout with fidelity greater than 95%. These results showcase a solid-state platform based on single coherent rare-earth ions for the future quantum internet.

**Main text:**

The distribution of entanglement over long distances using optical quantum networks is an intriguing macroscopic quantum phenomenon with applications in quantum systems for advanced computing and secure communication (*1, 2*). Solid-state emitters coupled to photonic resonators (*3*) are promising candidates for implementing quantum light-matter interfaces necessary for scalable quantum networks. A variety of systems have been investigated for this purpose, including quantum dots and defects in diamond or silicon carbide (*4–8*). So far, the ability to scale up these systems has remained elusive and motivates the development of alternative platforms. A central challenge is identifying emitters that exhibit coherent optical and spin transitions while coupled to photonic cavities that enhance the optical transitions and

channel emission into optical fibers. Ensembles of rare-earth ions (REIs) in crystals are known to possess highly coherent 4f-4f optical and spin transitions (*9*, *10*), but only recently have single REIs been isolated (*11*, *12*) and coupled to nanocavities (*13*, *14*). The crucial next steps toward using single REIs for quantum networks are demonstrating long spin coherence and single-shot readout in photonic resonators.

Here we demonstrate spin initialization, coherent optical and spin manipulation, and high-fidelity single-shot optical readout of the hyperfine spin state of single $^{171}\text{Yb}^{3+}$ ions coupled to a nanophotonic cavity fabricated in an yttrium orthovanadate (YVO) host crystal. The relevant energy level structure of $^{171}\text{Yb}^{3+}$ in YVO is shown in Fig. 1A (see Fig. S2 and Ref. (*15*) for additional details). $^{171}\text{Yb}^{3+}$ directly substitutes for $Y^{3+}$ in a site that has non-polar symmetry ($D_{2d}$), which reduces the sensitivity to electric field fluctuations that can cause optical decoherence. At zero applied magnetic field, the hyperfine interaction partially lifts the degeneracy of the ground state $^2F_{7/2}(0)$ leading to coupled electron-nuclear spin states of the form

$$|0\rangle_g = \frac{|\downarrow\Uparrow\rangle - |\uparrow\Downarrow\rangle}{\sqrt{2}}, |1\rangle_g = \frac{|\downarrow\Uparrow\rangle + |\uparrow\Downarrow\rangle}{\sqrt{2}}, \text{ and } |aux\rangle_g = |\uparrow\Uparrow\rangle, |\downarrow\Downarrow\rangle.$$

Here we denote the electron spin as $|\uparrow\rangle = \left|S_z = \frac{1}{2}\right\rangle, |\downarrow\rangle = \left|S_z = -\frac{1}{2}\right\rangle$ and the nuclear spin as $|\Uparrow\rangle = \left|I_z = \frac{1}{2}\right\rangle, |\Downarrow\rangle = \left|I_z = -\frac{1}{2}\right\rangle$. We use states $|0\rangle_g$ and $|1\rangle_g$, which are separated by ~675 MHz, to form the spin qubit. The $|0\rangle_g$ and $|1\rangle_g$ states have zero net magnetic moment and as a result the $|0\rangle_g \rightarrow |1\rangle_g$ transition is first-order insensitive to magnetic fluctuations that induce decoherence (*10*). The $|0\rangle_g \rightarrow |1\rangle_g$ transition retains the strength of the electron spin transition, which enables fast and efficient microwave manipulation.

A typical experimental sequence with spin initialization, control, and readout is shown in Fig. 1B. The $^{171}\text{Yb}^{3+}$ ions are coupled to a photonic crystal cavity with small mode volume ~$1(\lambda/n_{YVO})^3$ and large quality factor $(1 \times 10^4)$ (Fig 1C, D. See SI 1.1). This enhances the emission rate, collection efficiency, and cyclicity of the optical transitions *A* and *E* via the Purcell effect (*16*). The qubit is initialized into $|0\rangle_g$ by optical and microwave pumping on *F*, *A*, and $f_e$ to empty $|aux\rangle_g$ and $|1\rangle_g$, followed by cavity-enhanced decay into $|0\rangle_g$ via *E* (Fig. 1A). A subsequent microwave $\pi$ pulse applied on $f_g$ optionally initializes the ion into $|1\rangle_g$. The $|1\rangle_g$

state population is read out by exciting on $A$ and collecting the resulting ion fluorescence. Measurements are performed in a cryostat at 40 mK unless mentioned otherwise (See SI 6.5 for discussion of sample temperature). The ions are optically addressed using two frequency-stabilized continuous-wave lasers, while a microwave coplanar waveguide allows for driving of the spin transitions (Fig. 1E).

The YVO material used has a ~20 ppb residual concentration of $^{171}$Yb that is distributed over an optical inhomogeneous linewidth of ~200 MHz in the device due to variations in the local crystalline environment. This enables frequency isolation of single ions via pulsed resonant photoluminescence excitation (PLE) spectroscopy on transition $A$. The PLE scan in Fig. 2A shows peaks in fluorescence that are confirmed to originate from single ions by measuring the pulse-wise second-order photon correlation of the resonant emission (Fig. 2B). For ion X (marked in Fig. 2A), $g^2[0] = 0.147 \pm 0.011$. The observed bunching behavior for $t > 0$ is expected for a multi-level system with long-lived shelving states (See SI 2.3). An optical lifetime of $T_1 = 2.27\ \mu s$ is measured for ion X (Fig. 2C), which is a reduction from the bulk lifetime (267 $\mu s$) by $\beta F_p = 117$ (with $\beta = 0.35$ the branching ratio for emission via $A$) and corresponds to a single-photon coupling rate of $g = 2\pi \times 23$ MHz. Similar measurements were performed on ion marked as Y in Fig. 2A (See SI 2).

The cavity-enhanced optical transitions enable coherent optical control and efficient spin initialization (Fig. S7). Measurements of the resonant PL with varying excitation pulse length show optical Rabi oscillations (Fig. 2D), enabling calibration of $\pi$ and $\pi/2$ pulses for optical control and spin readout. An optical Ramsey measurement (Fig. 2E) gives a dephasing time of $T^*_{2,o} = 370$ ns, a factor of 12 shorter than the lifetime-limited $T_2 = 2T_1$. Further optical echo measurements give $T_{2,o} = 4.1\ \mu s$ (Fig. S8), which implies that $T^*_{2,o}$ is limited by quasi-static fluctuations of the transition frequency. We can extend the $T^*_{2,o}$ beyond 1 $\mu s$ by using post-selection to ensure the ion is on resonance with the readout sequence (Fig. S9). We measure the long-term stability, or equivalently the spectral diffusion, of the ion using PLE readout over 6 hours (Fig. 2F) and observe a narrow integrated linewidth (FWHM) of 1.4 MHz.

We use this optical initialization and detection to demonstrate coherent spin manipulation by driving Rabi oscillations on the $|0\rangle_g \to |1\rangle_g$ qubit transition (Fig. 3A). We perform a spin Ramsey measurement to extract a spin dephasing time of $T^*_{2,s} = 8.2\ \mu s$ (Fig. 3B). The spin coherence is further extended using dynamical decoupling sequences (*17*) to suppress quasi-

static contributions to dephasing. Fig. 3C shows the resulting coherence decay for increasing numbers of $\pi$ pulses using a Carr-Purcell-Meiboom-Gill (CPMG) sequence (Fig. 3C inset). For a single $\pi$ pulse, or spin echo sequence, we observe non-exponential behavior characteristic of a spin coupled to a slowly-fluctuating dipolar spin-bath (*18*, *19*) with $T_{2,s} = 43.5$ $\mu s$. This is further evidenced by a measurement of the coherence time with *N*, the number of $\pi$ pulses, which scales as $N^{0.70\pm0.01}$ (Fig. S9). CPMG scans taken with finer temporal resolution reveal periodic collapse and revivals of coherence indicative of coupling to nearby nuclear spins (Fig. S10) that could potentially be used as local quantum registers.

We explore the limits of the spin coherence time by increasing the number of rephasing pulses with a fixed pulse separation of 5.74 $\mu s$ to avoid unwanted interactions with the nuclear spin bath (*20*). This enables extension of the CPMG coherence time to 30 ms (Fig 3D). While the CPMG sequence does not allow for preservation of arbitrary quantum states, we also demonstrate coherence times longer than 4 ms using an XY-8 sequence (*17*) (Fig. S14) suitable for use in long-range quantum networks (*21*). The measured qubit lifetime of 54 ms (Fig. S15) indicates that the observed coherences are approaching the lifetime limit. We repeated these measurements at cryostat temperatures up to 1.2 K (Fig. 3D) and observed minimal changes in the spin coherence and lifetime, providing evidence that they are not limited by spin-lattice relaxation (See SI 6.5).

To harness this long spin coherence lifetime for quantum networks, it is essential to read out the qubit state in a single measurement. We achieve this with the scheme shown in Fig. 4A, which consists of two consecutive optical read periods on transition *A* separated by a microwave $\pi$ pulse to invert the qubit population. This scheme was designed considering that in this device direct resonant PL readout of the qubit state can only be performed using a series of optical $\pi$ pulses on transition *A* (See SI 2.2). The Purcell-enhanced cyclicity of transition *A* ($\beta_\parallel > 99.6\%$, see Fig. S6) allows for multiple photon emitting cycles before the ion is optically pumped out of the qubit subspace into $|aux\rangle_g$ ($|0\rangle_e \to |0\rangle_g$ is forbidden at zero-field). Fig. 4B shows the measured photon count distribution in the two readout sequences for the ion initialized in $|0\rangle_g$ (blue) or $|1\rangle_g$ (red). We assign the ion to $|1\rangle_g$ if we measure $\geq 1$ photons during the first readout sequence and 0 photons during the second readout sequence, and vice versa for $|0\rangle_g$. This discriminates between $|0\rangle_g$ and $|aux\rangle_g$ to ensure that the ion was in the qubit subspace during

the measurement. By implementing this scheme, we achieve an average readout fidelity of 95.3% (Fig. 4C).

These measurements showcase single $^{171}$Yb$^{3+}$ in YVO as a promising system for solid state quantum networking technologies. The measured spin coherence times correspond to light propagation for thousands of kilometers in optical fibers, which is necessary for long-distance quantum networks. Furthermore, the preservation of this coherence lifetime at temperatures of 1.2 K is promising for developing a viable technology with economical $^4$He cryogenics. To generate spin-spin entanglement with the current optical dephasing times will require a post-selection protocol similar to what has already been developed for other quantum networks (*22*) (Fig. S9). While the source of the optical dephasing is still under investigation, it will likely be improved in higher purity samples (See SI 6.5). While not explored here, the high magnetic field regime offers the possibility of longer optical and spin coherence times at the expense of weaker spin transition strengths (*15*). The crucial next steps are increasing the cavity Q/V by an order of magnitude and optimizing the collection efficiency, which should enable high rates of indistinguishable photon emission. This could be achieved with the current device architecture, or by using a hybrid platform where cavities are fabricated in a high-index material like GaAs and bonded to the YVO substrate. Multi-qubit gates necessary to establish entanglement on larger scale networks could be performed using the interaction with neighboring vanadium atoms or other REIs (*23*). The technology demonstrated here with single REI qubits complements other capabilities that could potentially be realized with $^{171}$Yb$^{3+}$:YVO, including quantum memories (*24*) for synchronizing photon traffic and quantum transducers (*25*) for coupling to qubits operating at microwave frequencies, thus pointing to a unified platform for the future quantum internet.

Note: While completing this work, we became aware of a related publication on single erbium ions showing results related to those presented in Fig. 4 (*26*).

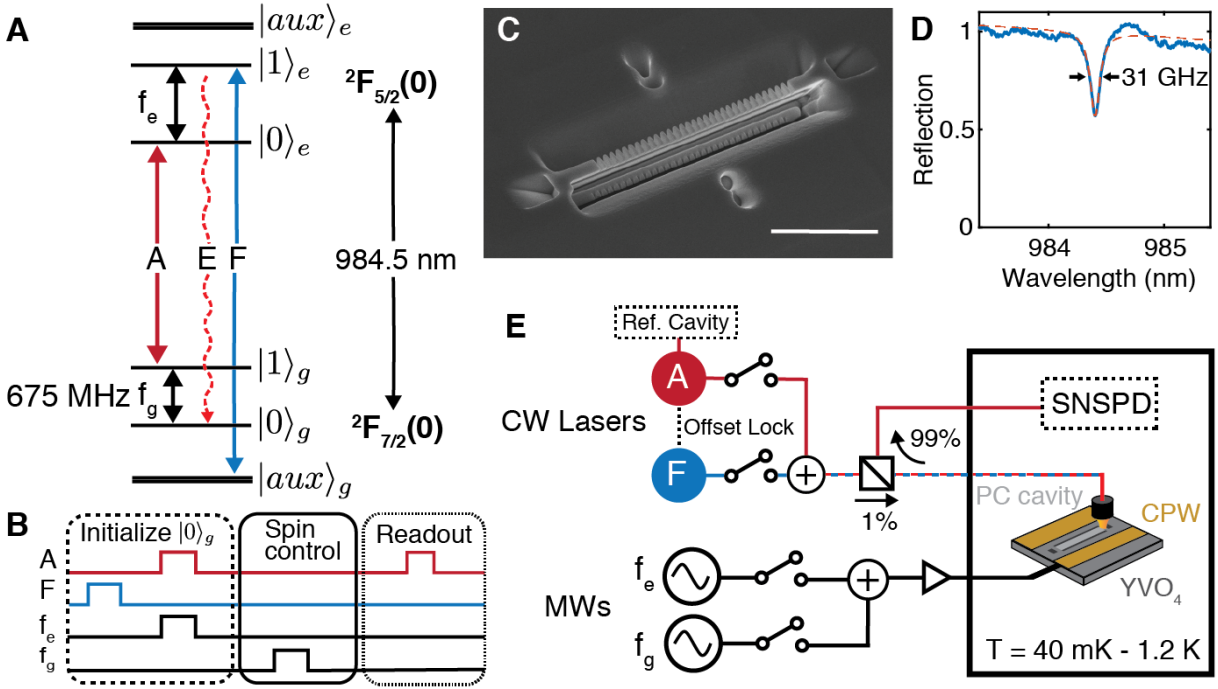

**Figure 1**: Experimental platform. A) Zero-field energy level structure of $^{171}\text{Yb}^{3+}$:YVO. States $|0\rangle_g$ and $|1\rangle_g$ form the spin qubit. Red transitions *A* and *E* are coupled (co-polarized) to the cavity, while the blue transition *F* is cross-polarized to the cavity mode. B) Typical experimental sequence used to initialize the ion into $|0\rangle_g$, manipulate the qubit, and optically read out the spin state. C) Scanning electron microscope image of a photonic crystal cavity fabricated in YVO (Scale bar = 10 $\mu m$). D) Reflection spectrum of the cavity. E) Schematic of experimental setup. The optical transitions are addressed using pulses generated from two frequency-stabilized lasers. The qubit is directly manipulated using microwave (MW) control pulses using a coplanar waveguide (CPW) next to the photonic crystal (PC) cavity, which is mounted in a dilution refrigerator. Light collected from the cavity is detected using a superconducting nanowire single photon detector (SNSPD).

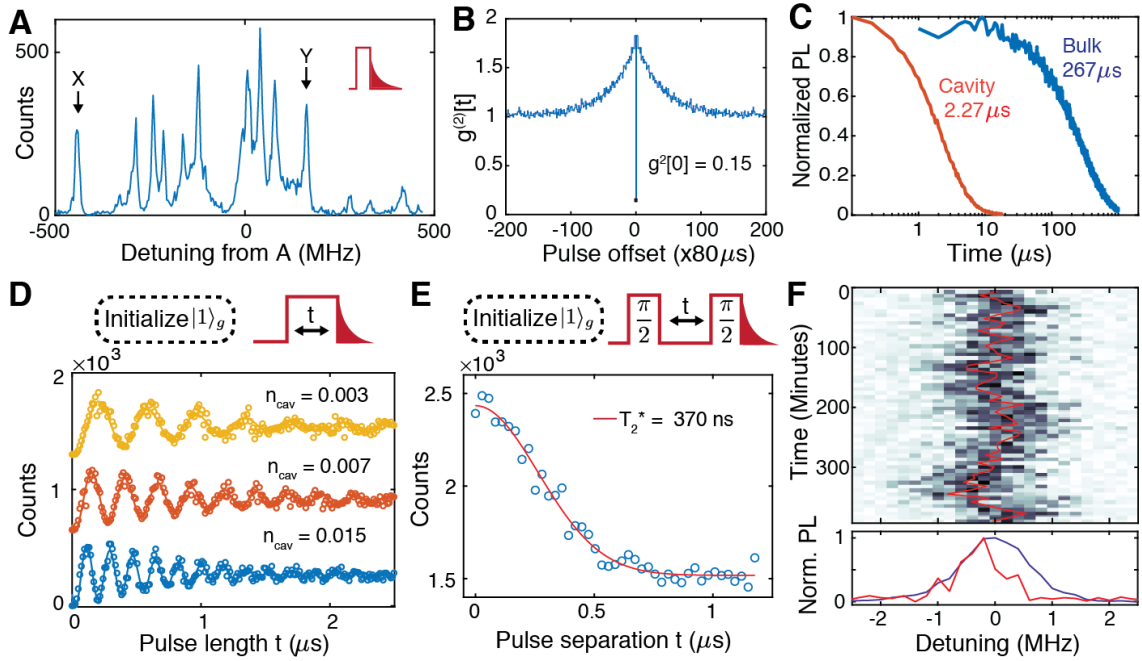

**Figure 2**: Optical detection and coherent optical manipulation of single $^{171}$Yb$^{3+}$ ions. A) Photoluminescence excitation (PLE) spectrum showing resolved peaks corresponding to single $^{171}$Yb$^{3+}$ ions. B) Pulsed autocorrelation measurement on ion X with $g^2[0] = 0.147 \pm 0.011$. C) Normalized photoluminescence emission from ion X coupled to the cavity (red) compared to typical photoluminescence from ions in bulk crystal (blue) showing lifetime reduction of ~120. D) Optical Rabi oscillations on transition $A$ after initialization into state $|1\rangle_g$ (top). The resulting fluorescence is plotted versus pulse length for different average cavity photon number n$_{cav}$. Plots are offset for clarity. E) Optical Ramsey measurement on transition $A$, indicating a dephasing time $T_2^* = 370$ ns. F) Measurement of spectral diffusion over six hours in which we repeatedly measure the transition frequency using PLE. Bottom plot shows typical scan (red) and sum of counts acquired during all scans (black), which is fit to a Gaussian with FWHM of 1.4 MHz.

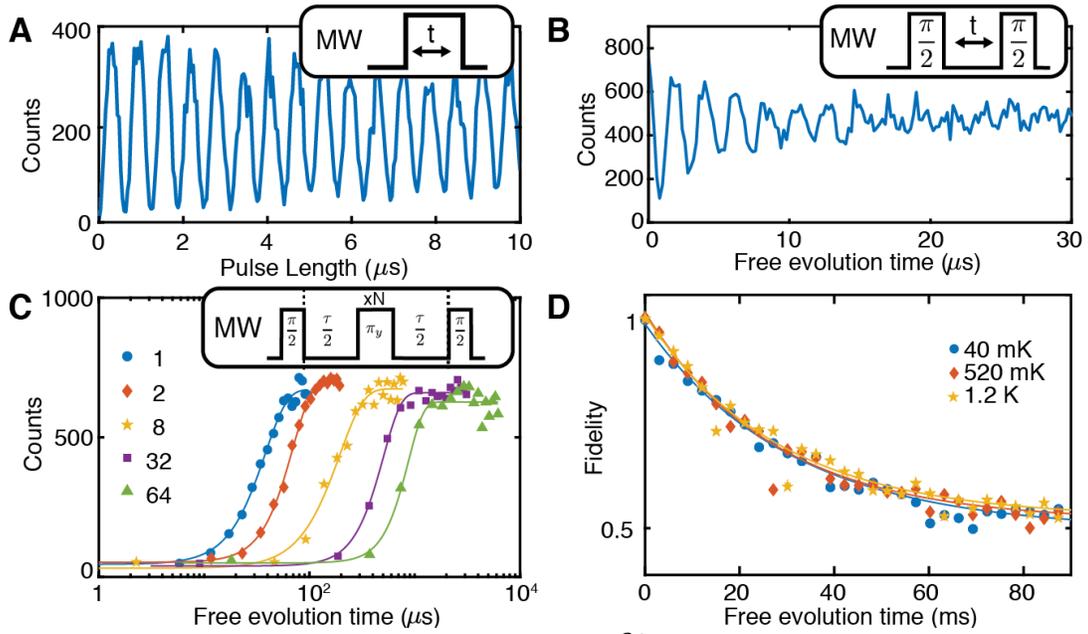

**Figure 3**: Coherent spin state control of a single $^{171}$Yb$^{3+}$ ion. A) Typical Rabi oscillations on the $|0\rangle_g \to |1\rangle_g$ microwave transition. B) Ramsey measurement on qubit transition (inset) that gives $T_{2,s}^* = 8.2$ μs. The excitation is detuned by 400 kHz to give rise to oscillations on the free-induction decay. C) Measurement of CPMG spin-coherence with increasing number of rephasing pulses. D) Measurement of spin coherence times up to 30 ms at temperatures up to 1.2 K using CPMG with fixed pulse separation and increasing numbers of rephasing pulses.

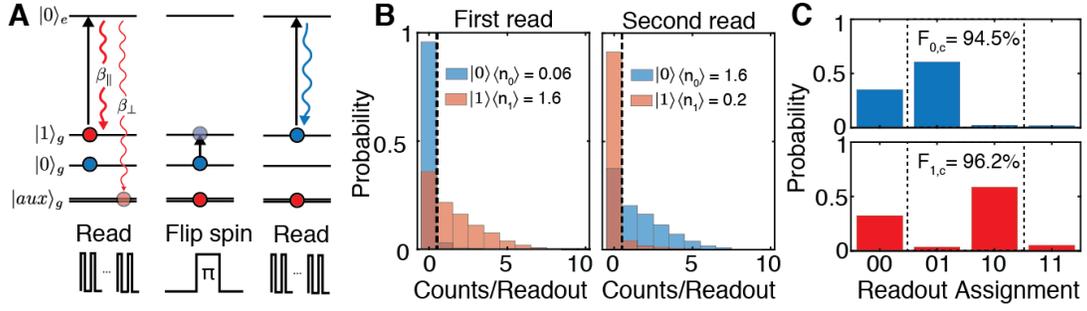

**Figure 4**: Single-shot readout (SSRO) of single $^{171}$Yb$^{3+}$ spin state. A) Scheme for SSRO. The ion is repeatedly excited using optical $\pi$ pulses on transition $A$ and the resulting fluorescence is collected. An ion in state $|1\rangle_g$ (red) will return to $|1\rangle_g$ with branching ratio $\beta_\parallel$ before eventually being pumped to $|aux\rangle_g$, while the ion in $|0\rangle_g$ (blue) will be largely unaffected by the readout. A microwave $\pi$ pulse is then applied to the spin transition to invert the population of $|0\rangle_g$ and $|1\rangle_g$ and the ion is optically read out again. The state of the ion is assigned based on the number of photons detected in the first and second read sequences. B) Photon-count distributions for the first and second read sequence used in the SSRO protocol with the ion initially prepared in $|0\rangle_g$ (blue) or $|1\rangle_g$ (red). The dashed line shows the state assignment threshold (1 photon). C) Distribution of assigned states $ij$ for dual readout scheme with the ion initially prepared in $|0\rangle_g$ (blue) or $|1\rangle_g$ (red), where $i(j)$ is the state assigned on the first (second) read. By conditioning the ion on the detection of 01 or 10, we obtain an average fidelity of 95.3%.

# References

1. H. J. Kimble, The quantum internet. *Nature*, **453**, 1023–1030 (2008).
2. S. Wehner, D. Elkouss, R. Hanson, Quantum internet: A vision for the road ahead. *Science,* **362** (2018)
3. D. D. Awschalom, R. Hanson, J. Wrachtrup, B. B. Zhou, Quantum technologies with optically interfaced solid-state spins. *Nat. Photonics*. **12**, 516–527 (2018).
4. B. Hensen *et al.*, Loophole-free Bell inequality violation using electron spins separated by 1.3 kilometres. *Nature*, **526**, 682–686 (2015).
5. W. F. Koehl, B. B. Buckley, F. J. Heremans, G. Calusine, D. D. Awschalom, Room temperature coherent control of defect spin qubits in silicon carbide. *Nature*, **479**, 84 (2011).
6. S. Sun, H. Kim, Z. Luo, G. S. Solomon, E. Waks, A single-photon switch and transistor enabled by a solid-state quantum memory. *Science,* **361**, 57–60 (2018).
7. A. Sipahigil *et al.*, An integrated diamond nanophotonics platform for quantum-optical networks. *Science,* **354**, 847–850 (2016).
8. D. D. Sukachev *et al.*, Silicon-Vacancy Spin Qubit in Diamond : A Quantum Memory Exceeding 10 ms with Single-Shot State Readout. *Phys. Rev. Lett.* **223602**, 1–6 (2017).
9. M. Zhong *et al.*, Optically addressable nuclear spins in a solid with a six-hour coherence time. *Nature*. **517**, 177–180 (2015).
10. A. Ortu *et al.*, Simultaneous coherence enhancement of optical and microwave transitions in solid-state electronic spins. *Nat. Mater.* **17**, 671–675 (2018).
11. R. Kolesov *et al.*, Optical detection of a single rare-earth ion in a crystal. *Nat. Commun.* **3**, 1029 (2012).
12. T. Utikal *et al.*, Spectroscopic detection and state preparation of a single praseodymium ion in a crystal. *Nat. Commun.* **5**, 3627 (2014).
13. T. Zhong *et al.*, Optically Addressing Single Rare-Earth Ions in a Nanophotonic Cavity. *Phys. Rev. Lett.* **121**, 183603 (2018).
14. A. M. Dibos, M. Raha, C. M. Phenicie, J. D. Thompson, Atomic Source of Single Photons in the Telecom Band. *Phys. Rev. Lett.* **120**, 243601 (2018).
15. J. M. Kindem *et al.*, Characterization of $^{171}$Yb:YVO$_4$ for photonic quantum technologies. *Phys. Rev. B*. **80**, 1–10 (2018).
16. E. M. Purcell, Spontaneous emission probabilities at radio frequencies. *Phys. Rev.* **69**, 681 (1946).
17. D. Suter, G. A. Álvarez, Colloquium: Protecting quantum information against environmental noise. *Rev. Mod. Phys.* **88**, 1–23 (2016).
18. G. De Lange, Z. Wang, S. Riste, V. V. Dobrovitski, R. Hanson, Universal Dynamical Decoupling of a Single Solid-State Spin from a Spin Bath. *Science*. **330**, 60–64 (2010).
19. J. R. Klauder, P. W. Anderson, Spectral Diffusion Decay in Spin Resonance Experiments. *Phys. Rev.* **125** (1961)
20. M. H. Abobeih *et al.*, One-second coherence for a single electron spin coupled to a multi-qubit nuclear-spin environment. *Nat. Commun.* **9**, 1–8 (2018).
21. P. C. Humphreys *et al.*, Deterministic delivery of remote entanglement on a quantum network. *Nature*. **558**, 268–273 (2018).
22. H. Bernien *et al.*, Heralded entanglement between solid-state qubits separated by three metres. *Nature*. **497**, 86–90 (2013).
23. M. Zhong, R. L. Ahlefeldt, M. J. Sellars, Quantum information processing using frozen

   core $Y^{3+}$ spins in Eu:$Y_2SiO_5$. *New J. Phys.* **21**, 033019 (2019).
24. T. Zhong *et al.*, Nanophotonic rare-earth quantum memory with optically controlled retrieval. *Science,* **357**, 1392–1395 (2017).
25. L. A. Williamson, Y.-H. Chen, J. J. Longdell, Magneto-Optic Modulator with Unit Quantum Efficiency. *Phys. Rev. Lett.* **113**, 203601 (2014).
26. Mouktik Raha, Songtao Chen, Christopher M. Phenicie, Salim Ourari, Alan M. Dibos, Jeff D. Thompson, Optical quantum nondemolition measurement of a solid-state spin without a cycling transition. *arXiv* (2019), 1907.09992

 
\setcounter{page}{1}
\maketitle

\tableofcontents
\listoffigures

\section{Experimental setup}
\subsection{Nanophotonic cavity in YVO}
Nanophotonic cavities are fabricated directly in a yttrium orthovanadate (YVO) crystal using focused-ion-beam (FIB) milling (Fig.~1C). Periodic trenches are made in a triangular nanobeam to form a photonic band gap with the spacing of these cuts tapered in the middle to form the defect required for the optical cavity mode. The reflectivity of one side of the cavity is lowered by reducing the number of photonic crystal lattice periods to allow for more efficient coupling into the collection path. Light is coupled into and out of these devices via total internal reflection using 45 degree couplers fabricated on both sides of the device. Further details on design and fabrication of nanocavities in YVO can be found in Ref. \cite{TZhong2016}. 

Devices are fabricated in a $c$-cut sample of YVO with the $E$-field of the fundamental TM mode aligned with the stronger optical dipole of Yb:YVO, which is polarized along the crystal $c$-axis. The device used here has an energy decay rate of $\kappa = 2 \pi \times 30.7$ GHz ($Q\sim 1 \times 10^4$). The mode volume extracted from FDTD simulations is $V = 0.095 \ \mathrm{\mu m^3}  \approx 1 (\lambda/n)^3$, where $n=2.17$ is the refractive index of YVO for $E \parallel c$. The coupling rate from free-space into the nanobeam waveguide is determined to be $\sim 24\%$ by direct measurement of the reflection from the device off resonance. The coupling rate of the input mirror of the cavity, $\kin$, is extracted from the cavity reflection spectrum (Fig.~1D) to be $\kin/\kappa \approx 0.14$. The cavity is determined to be undercoupled by measuring the phase response using a polarization interferometer \cite{Tiecke2014}. 

The sample used for this work is cut and polished from a boule of YVO grown by Gamdan Optics. While nominally undoped in the growth process, the crystal contained residual concentrations of rare-earth ions. From optical absorption measurements in bulk crystals and glow discharge mass spectrometry (GDMS, EAG laboratories), the total concentration of all Yb isotopes is estimated to be 0.14 ppm. Assuming natural isotopic abundance ($14.3\%$\ybion$\!\!$), this gives a \ybion concentration of $\sim20$ ppb, which corresponds to $\sim23$ \ybion ions within the cavity mode volume.

\subsection{Detailed experimental setup}

\begin{figure}[hb]
    \centering
    \includegraphics[width=\columnwidth]{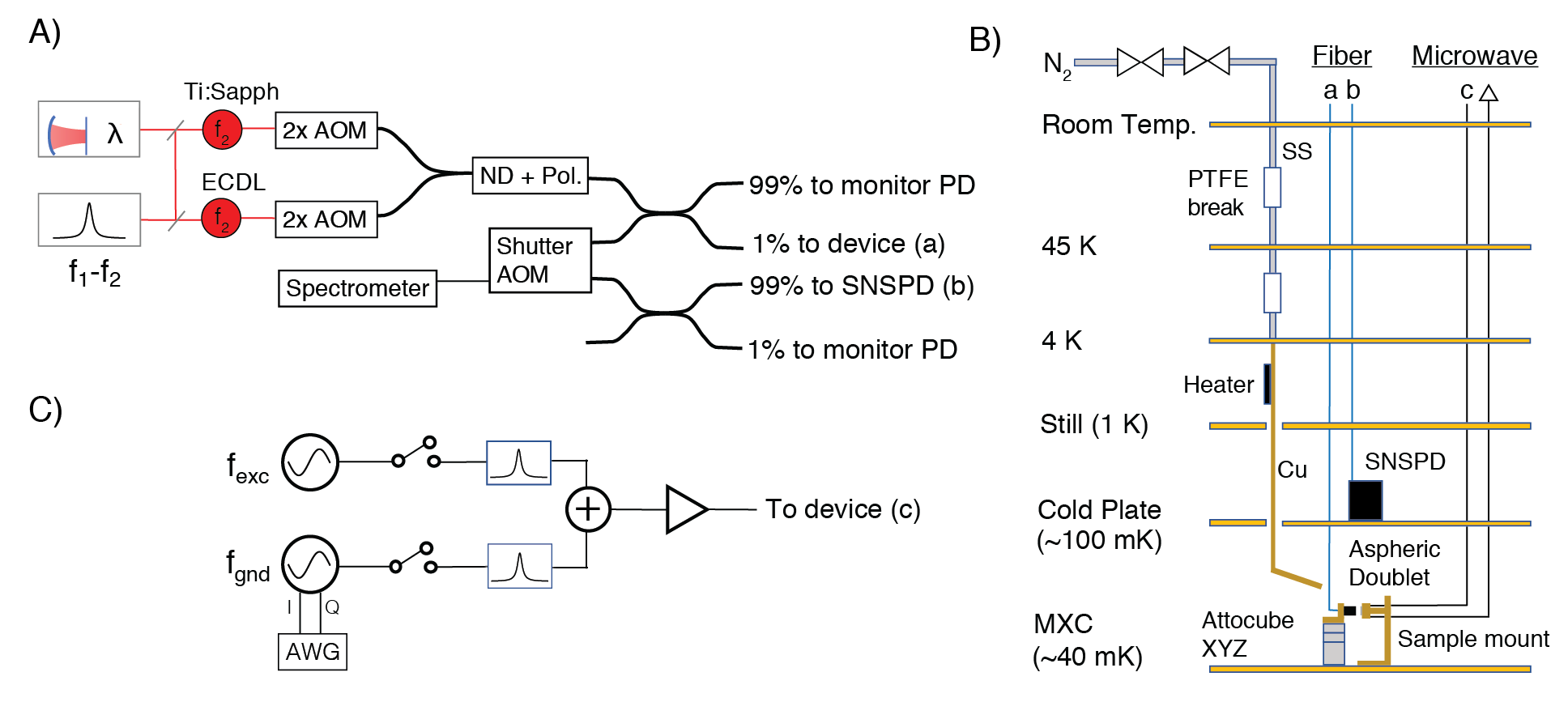}
    \caption[Experimental Setup]{Additional details of experimental setup. A) Optical network for optical manipulation and readout of single ions. Light from two single-mode lasers (Ti:Sapph and ECDL) are modulated by AOMs, coupled into fiber, and passed through neutral density (ND) filters and a polarization controller. Light is sent to the device in the dilution refrigerator using a 99/1 fiber splitter, with $1\%$ going to the device (a) and $99\%$ monitored on a photodiode. Light reflected from the device is directed to an acousto-optic modulator (AOM) before being detected using a $\mathrm{WSi_2}$ superconducting nanowire single photon detector (SNSPD) (b). The reflected signal can also be routed to a spectrometer for alignment and tuning of the device. A small portion of the Ti:Sapph is used to establish an offset-frequency lock to a Fabry-Perot reference cavity and is monitored by a wavemeter. The ECDL is held at a fixed frequency with respect to the Ti:Sapph using by measuring the beat note between the two lasers on a fast photodiode. B) Schematic of experiment inside the dilution refrigerator. Light from the optical network in (A) is coupled into the device via an aspheric doublet mounted on an Attocube XYZ stack. Tuning is accomplished by depositing $\mathrm{N_2}$ on the device as described in the text. C) Microwave setup for manipulating spin transitions. Two microwave frequency sources are sent through switches and narrow-band filters before being combined, amplified, and sent to the device in the dilution refrigerator (c). }
    \label{fig:experimentalsetup}
\end{figure}

\fig{experimentalsetup}A shows a schematic of the optical network used in these experiments. The ions in the cavity are optically addressed using two continuous-wave lasers. The state of the ions is read out on transition $A$ (Fig.~1A) using a Ti:Sapphire laser (M2 Solstis) and optical pumping on transition $F$ is performed using an external-cavity diode laser (ECDL, Toptica DLPro).

A small portion of the Ti:Sapph is picked off to enable locking to a high-finesse Fabry-Perot cavity (Stable Laser Systems) that serves as a stable long-term frequency reference. A fiber-based phase-modulator (EOSpace) imposes variable-frequency sidebands onto the light before the reference cavity and the first-order sideband is locked to the cavity using the standard Pound-Drever-Hall (PDH) technique. Scanning the frequency of this sideband enables quasi-continuous scans over a 3 GHz range while locked to a single longitudinal mode of the reference cavity. The wavelength of the laser is monitored using a wavemeter (Bristol Instruments) to reliably lock to the same longitudinal mode of the cavity. The ECDL is held at a fixed frequency with respect to the Ti:Sapph by measuring the beat note between the two lasers on a fast photodiode (Newport) and feeding back to the ECDL current and piezo control. 


Each laser is independently amplitude-modulated using two free-space acousto-optic modulators (AOMs) in double-pass configuration with total extinction of $\approx120$ dBm. The lasers are coupled into fiber and combined using a fiber-based beamsplitter. A set of variable attenuators and polarization controllers allow for further adjustment of the amplitude and polarization of the light that passes through a 99/1 fiber splitter before being directed to the device. Light reflected or emitted from the device is directed by the 99/1 splitter through an additional free-space AOM shutter before being detected by a $\mathrm{WSi_2}$ superconducting nanowire single photon detector (SNSPD) \cite{Marsili2013}. This detector has high efficiency ($\sim75 \%$) and low intrinsic dark counts ($<1$ Hz). The AOM shutter serves to protect the SNSPD from latching during preparation and readout pulses. All fiber connections are spliced when possible to minimize reflections and additional losses. The total system detection efficiency (probability of detecting a photon emitted by an ion in the cavity) is $\sim1\%$.

A broadband supercontinuum source (Fianiuum WhiteLase Micro) and a spectrometer (Princeton Instruments SP-2750, PIXIS 2KB eXcelon) are used for measuring the cavity reflection spectrum when aligning and tuning the device.

\fig{experimentalsetup}B shows a schematic of the experimental setup inside of a Bluefors LD250 dilution refrigerator. The device is held stationary on a copper sample mount on the mixing chamber (MXC) plate. Light is coupled into and out of the device from fiber using an aspheric doublet mounted on an XYZ piezo-stage (Attocube) that allows for optimization of this coupling at dilution fridge temperatures. 

Devices are tuned onto resonance with the ion transition of interest by nitrogen deposition \cite{Mosor2005}. To implement this in the dilution refrigerator, a gas tuning line is installed from room temperature down to the mixing chamber (\fig{experimentalsetup}B). This tuning line consists of stainless-steel (SS) tubing from room temperature to 4 K with the tubing thermalized at each stage and isolated between stages by a PTFE break. From 4 K to the mixing chamber plate, the tuning line consists of a free-hanging copper tube that is thermally isolated from the components below 4 K. The output of this line is directed onto the sample on the mixing chamber plate. A resistive heater attached to the tuning line near the 4 K stage enables warming of the line such that gas flows through the line without freezing and is deposited onto the device. Careful adjustment of the heater power allows for fine red-tuning of the cavity resonance at rates $< 0.1$ nm/minute. The cavities can be detuned by sublimation of the frozen nitrogen through optical heating of the device using $\sim100\; \mathrm{\mu W}$ of laser power resonant with the cavity mode.

\fig{experimentalsetup}C shows the setup for microwave control of single ions. Microwave tones to drive the ground and excited state transitions are generated using two signal generators (Stanford Research Systems SG380). The amplitude and phase of the pulses used on the ground state transition are controlled using IQ modulation driven by a fast function generator (Tektronix AWG5204). Both sources pass through a set of microwave switches (Minicircuits ZASWA-2-50DR+) that provide additional extinction before passing through narrow-band filters. The microwave tones are combined, amplified, and sent to the device in the dilution fridge. To ensure adequate microwave power at the device for these initial measurements, minimal attenuation is used on the input coaxial lines inside the fridge with a single 20 dB attenuator on the still plate and 0 dB attenuators on the other plates. 


A gold coplanar waveguide is fabricated next to the optical cavity to allow for microwave manipulation of the ions. The center strip of this waveguide is 60 $\mathrm{\mu m}$ wide with a spacing of 30 $\mathrm{\mu m}$ to the ground plane. The optical device sits inside this 30 $\mathrm{\mu m}$ gap. Launching microwaves through this waveguide gives rise to an oscillating magnetic field along the crystal $c$-axis, which enables driving of the desired transitions ($\ket{0}\rightarrow \ket{1}$) at zero-field. The YVO chip sits inside a microwave launch board (Rogers AD1000, fabricated by Hughes Circuits) with SMP connectors on both input and output. This launch board is wire-bonded to the chip with as many wirebonds as possible to give additional cooling through the surface.

Static magnetic fields are applied to the device inside the fridge using a set of homebuilt superconducting magnets made by winding superconducting wire (SC-T48B-M-0.254mm, Supercon Inc).

With the full experiment loaded, the temperature of the mixing chamber plate is $\sim40$ mK. Experiments are performed with the mixing chamber plate temperature up to 1.2 K to investigate temperature dependence of spin coherence and lifetime. Further measurements above this temperature have not been performed at this time as this leads to spurious dark counts and increased latching on the SNSPDs. Further measurements of the temperature of the device are presented in \sect{lifetimeandtemp}.

\section{Identifying single $\mathbf{{}^{171}Yb}$ ions}

\subsection{Energy structure of $\mathbf{{}^{171}Yb:YVO}$}
\label{sec:energystructure}
The $4f^{13}$ configuration of $\mathrm{Yb}^{3+}$ consists of two electronic multiplets, ${}^2F_{7/2}$ (ground state) and ${}^{2}F_{5/2}$ (excited state), that are split by the crystal field of \yvo \; into four and three Kramers doublets, respectively. The optical transition of interest is between the lowest energy doublets of the ground state and excited state (${}^2F_{7/2}(0) \rightarrow {}^{2}F_{5/2}(0)$), which occurs at approximately 984.5 nm for $\mathrm{Yb}^{3+}$ doped into \yvo. At cryogenic temperatures, the Kramers doublets can be treated as spin-$1/2$ systems and described using an effective spin Hamiltonian \cite{Abragam1970}. For isotopes with non-zero nuclear spin, the hyperfine interaction adds additional energy structure. The ${}^{171}\mathrm{Yb}$ isotope is unique among the rare-earth ions as the only stable Kramers ion with nuclear spin $I = 1/2$, which gives rise to the simplest possible level structure with both electron and nuclear spin. Additional discussion on the level-structure and effective spin Hamiltonian can be found in \cite{Kindem2018}.

\begin{figure}
\centering
    \includegraphics{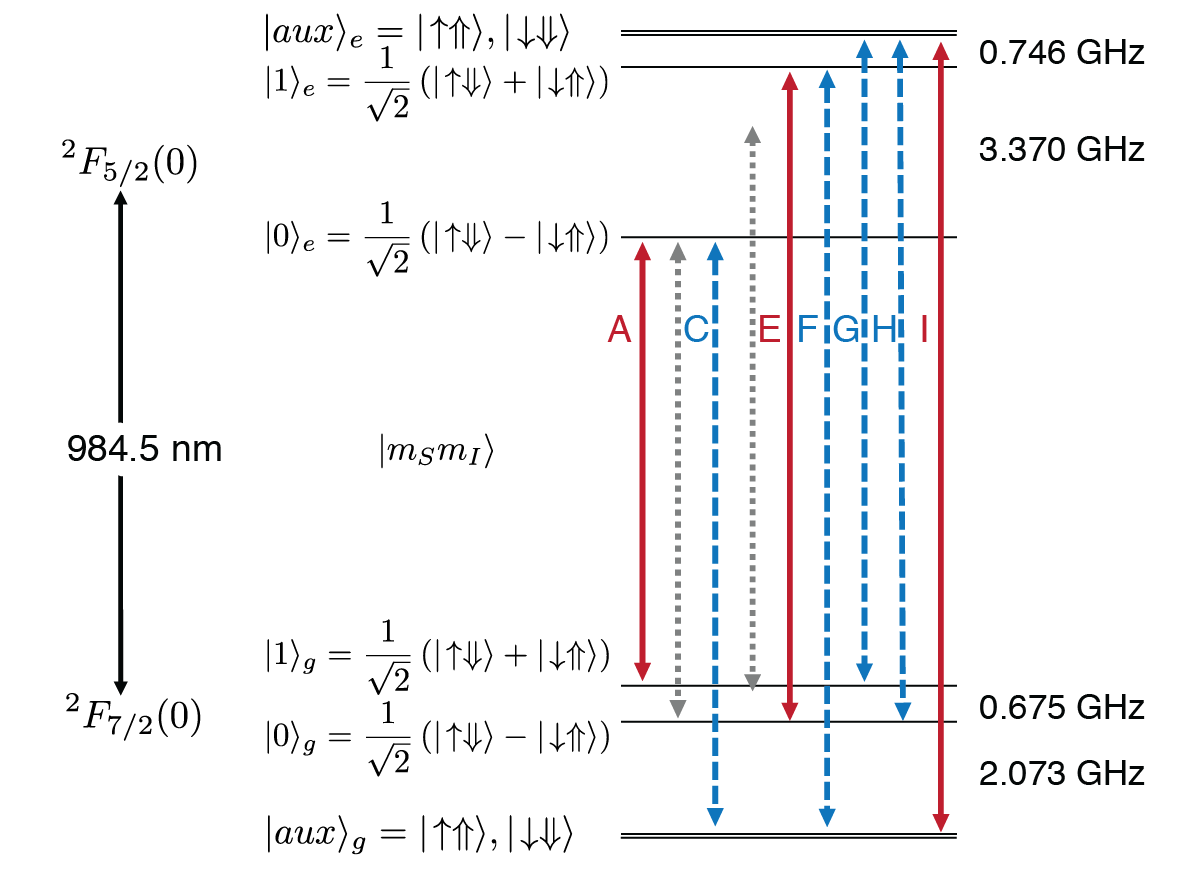}
    \caption[Detailed level structure]{Detailed level structure of \ybyvo at zero-field \cite{Kindem2018}. We denote the electron spin component of the state as $\ket{\uparrow} \equiv \ket{S_z = \frac{1}{2}}$, $\ket{\downarrow} \equiv \ket{S_z = - \frac{1}{2}}$  and the nuclear spin component as $\ket{\Uparrow} \equiv \ket{I_z = \frac{1}{2}}$, $\ket{\Downarrow} \equiv \ket{I_z = -\frac{1}{2}}$. States $\ket{0}_g$ and $\ket{1}_g$ form the spin qubit used in this work. States $\ket{\uparrow \Uparrow}$ and $\ket{\downarrow \Downarrow}$ are degenerate in the absence of strain and are referred to throughout the text as $\ket{aux}$. Optical transitions allowed for $E\parallel c$ are shown in red (solid), while transitions allowed for $E \perp c$ are shown in blue (dashed).}
    \label{fig:detailedlevelstructure}
\end{figure}

The energy level structure of the lowest crystal field levels of \ybyvo at zero-field is shown in \fig{detailedlevelstructure}. The red (blue) lines correspond to optical transitions allowed for light polarized parallel (perpendicular) to the $c$-axis of the crystal. The red transitions are the Purcell-enhanced transitions co-polarized with the cavity mode. The transitions $\ket{0}_g\rightarrow \ket{0}_e$ and $\ket{1}_e\rightarrow \ket{1}_g$ are forbidden at zero-field by symmetry. 

\subsection{PLE scans}

\begin{figure}
    \includegraphics[width=\columnwidth]{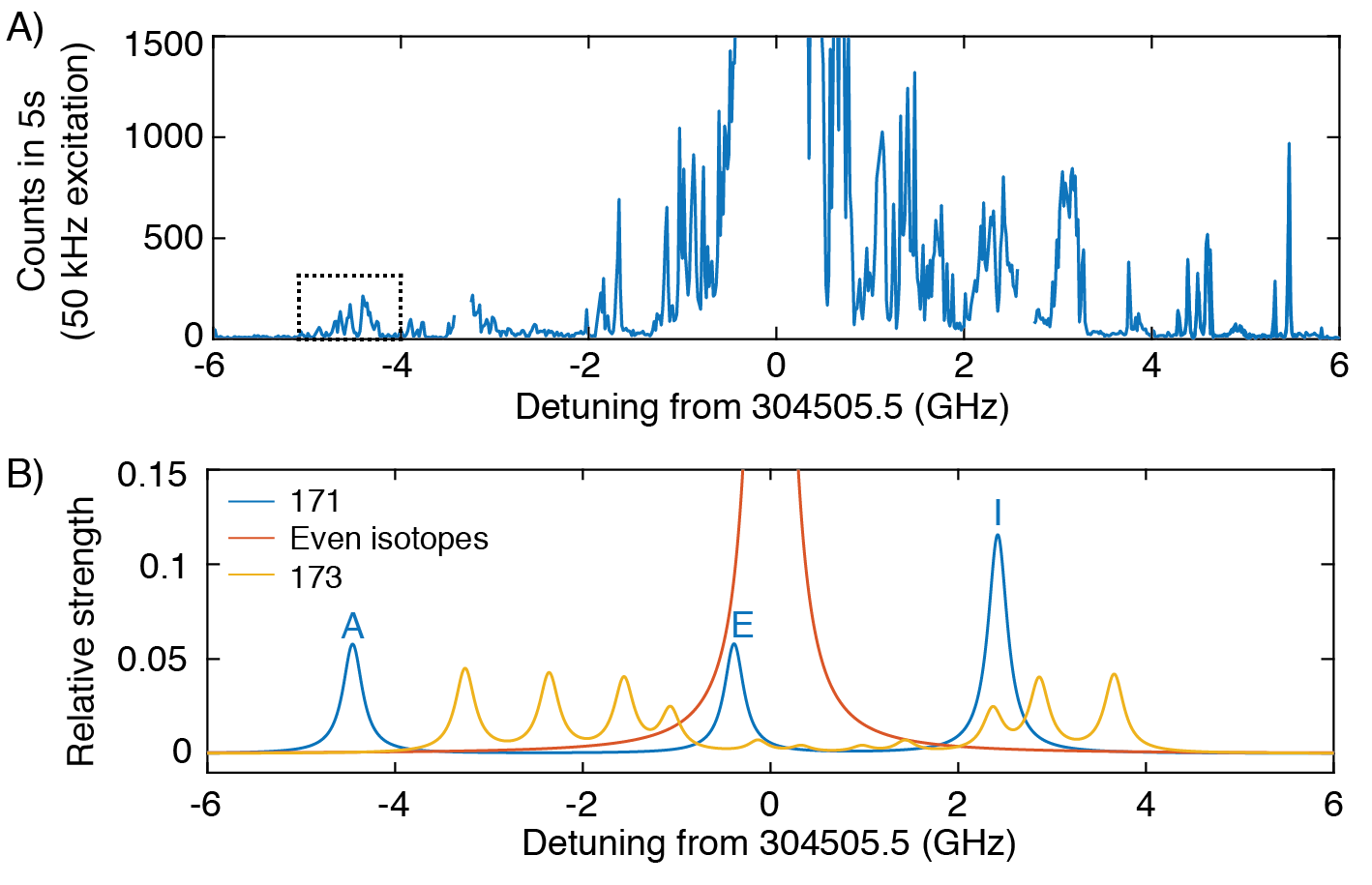}
    \caption[PLE scans and expected ion distribution]{A) Extended photoluminescence scan over 12 GHz centered around the optical transition of the zero-nuclear-spin isotope with zero applied magnetic field. The dashed box highlights the region scanned in the Fig.~2A of the main text. B) Predicted optical transition frequencies of the different Yb isotopes for $E \parallel c$ with transition strength scaled to natural abundance.}
    \label{fig:fulllinescans}
\end{figure}

Potential single ions are identified with pulsed resonant photoluminescence excitation (PLE) scans. \fig{fulllinescans}A shows an extended PLE line scan over a 12 GHz region around the center of the optical transition. Clusters of peaks in fluorescence correspond to the different isotopes of Yb, which have the expected transitions shown in \fig{fulllinescans}B. These PLE scans are taken with Rabi frequencies $> 10$ MHz to intentionally power-broaden the optical transitions of the ions and enable coarser and faster scans. As shown in \fig{fulllinescans}B, transition $A$ of Yb-171 does not spectrally overlap with optical transitions from the other isotopes, while the other cavity-coupled optical transition from the qubit subspace (transition $E$) overlaps with the inhomogeneous distribution of the zero-spin isotope. This makes it difficult to isolate and address single Yb-171 ions using transition $E$ without simultaneously exciting a large number of zero-spin ions. As a result, finer scans are performed around transition $A$ (Fig.~2A of the main text) to identify potential Yb-171 ions. 

\begin{figure}
    \includegraphics[width=\columnwidth]{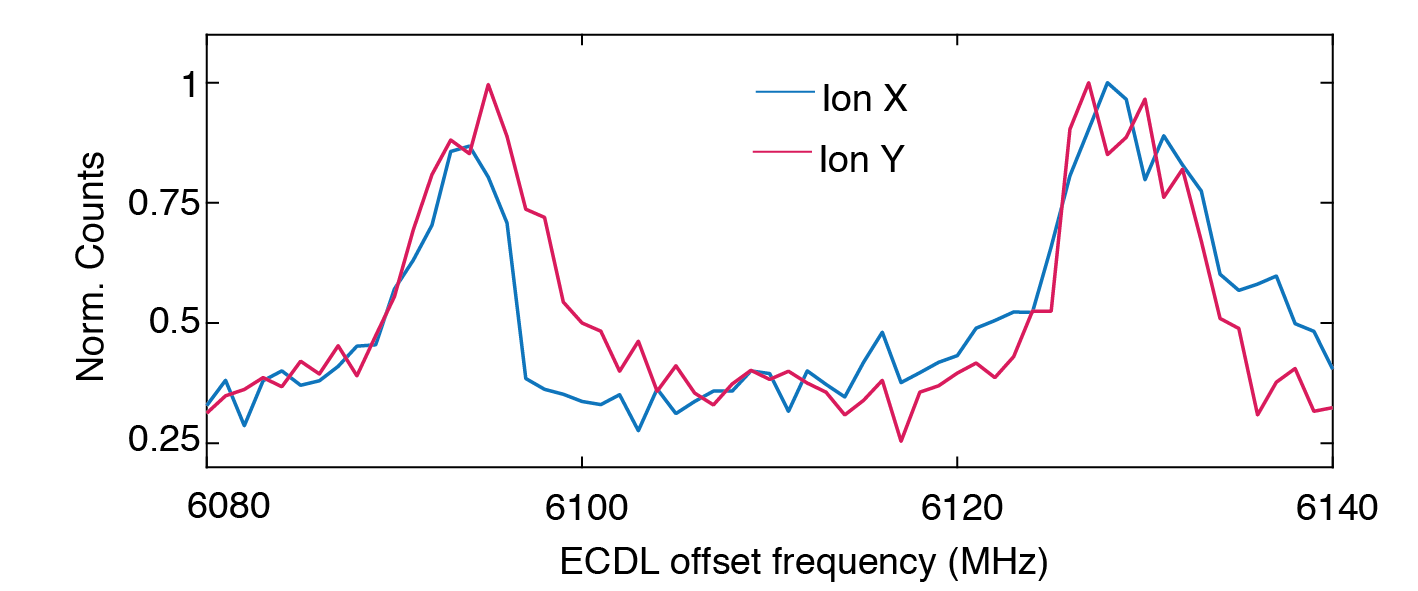}
    \caption[Identification of Yb-171 ions]{Offset frequency scans to verify potential Yb-171 ions identified in Fig.~2A. A readout pulse is tuned on resonance with an isolated peak and a second laser is used to perform optical pumping around transition $F$.}
    \label{fig:strainsplitting}
\end{figure}

To determine whether an isolated peak corresponds to a Yb-171 ion, we investigate the energy level structure using optical pumping. The readout laser is tuned on resonance with one of these peaks and a second laser is scanned across transition $F$. If an observed peak corresponds to the $A$ transition of a Yb-171 ion, the pump laser will move population into the qubit subspace as it comes into resonance with $F$ and result in an increase in counts after the readout pulse. \fig{strainsplitting} shows examples of these scans performed on the ions labeled as X and Y in Fig.~2A. A small splitting of transition $F$ is observed that is unexpected for the ion at zero magnetic field. Further investigations into the behavior of this splitting with applied magnetic field confirm that this is not due to a residual magnetic field at the ion. 

This splitting is attributed to the ions occupying strained or otherwise distorted sites in the crystal. Here, we isolate single ions by working in the tails of the inhomogeneous distribution arising from variations in the local environment within the crystal. As a result, we are in practice preferentially selecting for strained ions. A distortion of the local crystal lattice can reduce the site symmetry of the ion, which would lead to a breaking of the degeneracy of the lowest energy levels in the ground state (i.e. break the degeneracy of $\ket{aux}_g$) \cite{Ortu2017}. Further studies are necessary to understand the nature and cause of this strain and its consequences for the properties of the ion.  Depending on the resulting site symmetry of the ion, these ions could potentially now possess a DC Stark shift \cite{Macfarlane2007}. This would have negative implications for long-term optical spectral diffusion due to fluctuating electric fields, but would also open the door to tuning and stabilization of the optical transition through applied DC electric fields. 

\subsection{Verifying single ions}

Second-order intensity correlation measurements are performed to verify that an isolated peak corresponds to emission from a single ion. In Fig.~2B, $g^{(2)}[t]$ on ion X is measured by alternating between a single initialization pulse on the $C$ transition and a readout pulse on transition $A$. The pulsewise correlation is calculated on the counts observed after the excitation pulse on transition $A$. We note that because the detector deadtime is short compared to the excited state lifetime and photon rate, these measurements were performed using a single detector and by calculating a full autocorrelation.

The bunching behavior observed for $t>0$ is expected for a multi-level system with long-lived shelving states, where the amplitude of the bunching corresponds to the ratio of effective decay rates into and out of $\ket{1}_g$ \cite{Kitson1998}. The single initialization pulse is not sufficient in this case to completely initialize the ion into $\ket{1}_g$ before each readout, but was chosen to enable faster repetition of the experiment. We expect that this effect would not be observed with full initialization of the ion before each readout. This bunching could also be attributed to spectral diffusion or blinking \cite{Besombes2010}. Additional $g^{(2)}[t]$ measurements performed on ions identified as zero-spin isotopes at zero-field (i.e. no shelving levels) do not exhibit this bunching, providing further evidence that this is due to the multiple long-lived levels of Yb-171 at zero-field. 

\begin{figure}
    \includegraphics[width=\columnwidth]{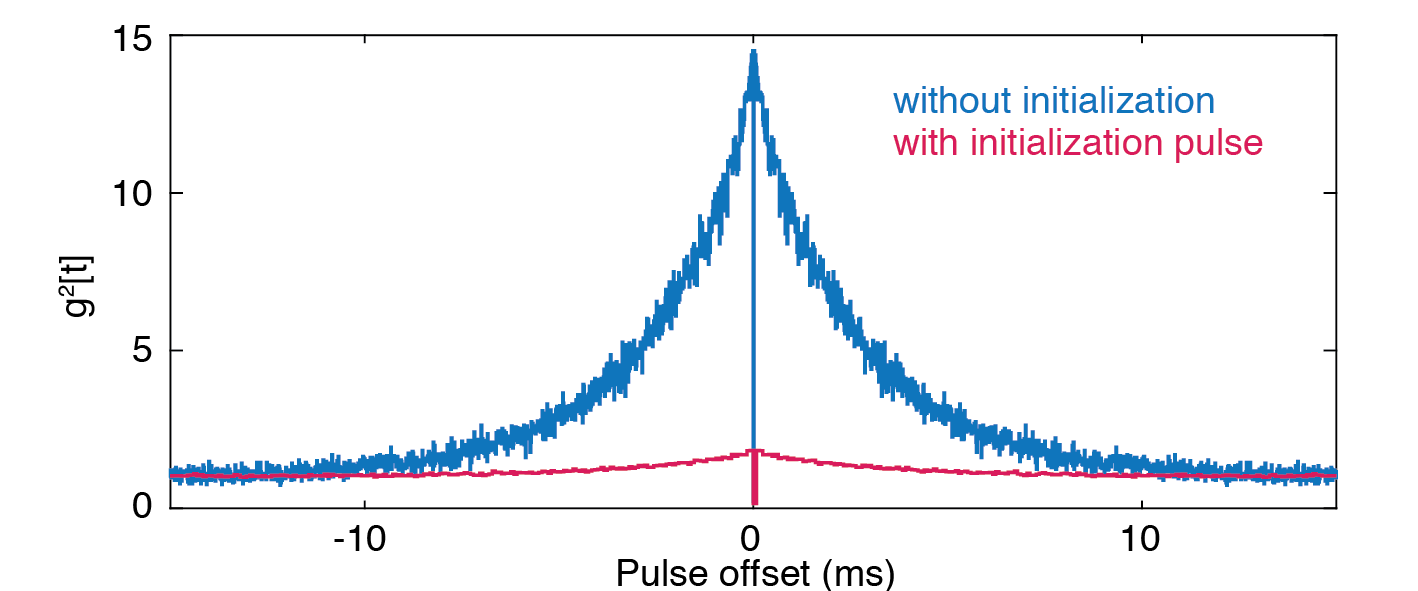}
    \caption[Additional $g^{(2)}$ measurements]{Measurement of pulsed second-order autocorrelation on ion X with a single initialization pulse on transition $C$ (red) and no initialization pulse (blue).}
    \label{fig:additionalg2}
\end{figure}

To show that the observed bunching behavior is related to a population effect, this measurement was repeated without any initialization into state $\ket{1}_g$ as shown in \fig{additionalg2}. In this case, the rate of pumping into $\ket{1}_g$ is reduced, while the effective pumping rate out of $\ket{1}_g$ is held constant. This leads to a drastically increased bunching behavior for $t > 0$ as shown.

Measurements performed on ion Y show similar behavior and give $g^{(2)}[0] =0.30\pm0.03$.  These results indicate that we have correctly identified X and Y as single \ybion ions.

\section{Purcell enhancement and optical branching ratio}
\subsection{Predicted Purcell enhancement}

The oscillator strength of the ${}^2F_{7/2}(0) \rightarrow {}^{2}F_{5/2}(0)$ transition for light polarized along $c$ (i.e. transitions $A$, $E$, $I$) was determined to be $f = 4.8 \times 10^{-6}$ from bulk absorption measurements \cite{Kindem2018}. Note that here we are using the real-cavity local field correction factor between absorption and oscillator strength \cite{Schuurmans2000}. The corresponding dipole moment of these transitions is $1.06 \times 10^{-31} \; \mathrm{C \cdot m}$, which gives an emission rate for $E \parallel c$ of $1/(763 \,\mathrm{\mu s})$. Using the bulk excited state lifetime of $267 \, \mathrm{\mu s}$, this gives a branching ratio for decay with $E \parallel c$ of $\beta_{\parallel}\sim 0.35$.

The optical decay rate of the atom in the nanophotonic cavity, $\gamma_{cav}$, is enhanced from its free space value $\gamma_0 = 1/(267 \, \mathrm{\mu s})$ by
\beq
\frac{\gamma_{cav}}{\gamma_0} = 1+ \frac{4 g^2 }{\kappa \gamma_0}
= 1 + \eta,
\eeq
where we have assumed that the cavity is resonant with the optical transition. Here, $\eta$ is referred to as the effective Purcell factor to distinguish from the enhancement of the cavity-coupled transition by $F_p$ and the resulting overall change in the lifetime determined by the branching ratio (i.e. $\eta = \beta_{\parallel} F_p$). The coupling between atom and cavity field is described by the single photon Rabi frequency, $2g$, where
\beq
g = \frac{\mu}{\hbar}\sqrt{\frac{\hbar \omega}{2 \epsilon_0 n^2 V}},
\label{eq:theory_g}
\eeq
$\mu$ is the transition dipole moment, $n$ is the refractive index of the medium, and $V$ is the optical mode volume of the cavity. The cavity energy decay rate is $\kappa = 2 \pi \times 30.7 \, \mathrm{GHz}$ (Fig.~1D). For simplicity, we assume that the ion is placed at the maximum of the cavity field and optimal polarization alignment between the cavity mode and the transition dipole.

For the system parameters presented above, the maximal expected coupling is $g_{max} = 2 \pi \times 25.5 \, \mathrm{MHz}$. The maximum effective Purcell enhancement in the cavity used here is then $\eta_{max} = 143$, which corresponds to a cavity lifetime of $1.87 \, \mathrm{\mu s}$. This is in reasonable agreement with the measured lifetime of ion X of $2.27 \, \mathrm{\mu s}$, which corresponds to an effective Purcell enhancement of $\beta F_p = 117$. The resulting cavity QED parameters for this system are $(g, \,\kappa,\, \gamma)$ = $2 \pi \times (22.9 \, \mathrm{MHz}, 30.7 \, \mathrm{GHz}, 596 \, \mathrm{Hz})$. Similar measurements on ion Y give a Purcell-enhanced lifetime of $2.3\,\mathrm{\mu s}$ indicating that ion X and ion Y are nearly identically coupled to the cavity.

In addition to enhancing the emission rate, the Purcell effect leads to preferential emission of photons into the cavity mode from which they can be more readily collected. The fraction of the ion emission into the cavity mode, $P_{cav}$, is given by the ratio of emission into the cavity to the total emission rate:
\beq
P_{cav} = \frac{\beta F_p}{1+ \beta F_p}.
\eeq
The measured Purcell enhancement corresponds to $P_{cav} = 99.1\%$. 

\subsection{Modification of branching ratio in cavity}
\label{sec:branchingratio}
The Purcell enhancement in this system improves the cyclicity of the optical transitions. In this context, cyclicity describes the probability that an excited ion will return to its original ground state upon emission of a photon. High cyclicity is essential for single-shot readout in which the qubit state is assigned based on the number of photons detected during repeated optical excitation of the ion.

We assume the ion starts in $\ket{1}_g$ and is excited to $\ket{0}_e$ on transition $A$.  Once in the excited state, the ion can decay via the 984.5 nm transition back to $\ket{1}_g$  with rate $\gamma_\parallel$ or to $\ket{aux}_g$ with $\gamma_\perp$. It can also decay back to the ground state through the other crystal field levels with rate $\gamma_{other}$. The total excited state decay rate $\gamma_0$ is then
\beq
\gamma_0 = \gamma_\parallel + \gamma_\perp + \gamma_{other}.
\eeq
For an ion in the bulk crystal, the overall branching ratio for decay via $A$ is $\beta_{\parallel} = \gamma_\parallel/\gamma_0 \approx 0.35$ \cite{Kindem2018}. The cavity enhances the emission rate for $E \parallel c$ by $1+ F_p$, which results in a cavity-enhanced branching ratio for this transition:
\begin{align}
\beta_{\parallel}^{cav} & = \frac{(1 + F_p) \beta_{\parallel} \gamma_0}{\gamma_{cav}} = \frac{(1 + F_p) \beta_{\parallel}}{1 + F_p \beta_{\parallel} }  \\
&= 1 - \of{1-\beta_\parallel}\frac{T_{1}^{cav}}{T_1^{bulk}}.
\end{align}

From the observed cavity lifetime of $T_1^{cav} = 2.3\, \mathrm{\mu s}$ and bulk lifetime of $T_1^{bulk} = 267 \, \mathrm{\mu s}$, we predict a branching ratio in the cavity of  $\beta_{\parallel}^{cav} \ge 0.994$. Here we have assumed that decay through the other crystal field levels will bring the ion to a different ground state to provide a lower bound to the expected branching ratio in the cavity.

\begin{figure}
    \includegraphics[width=\columnwidth]{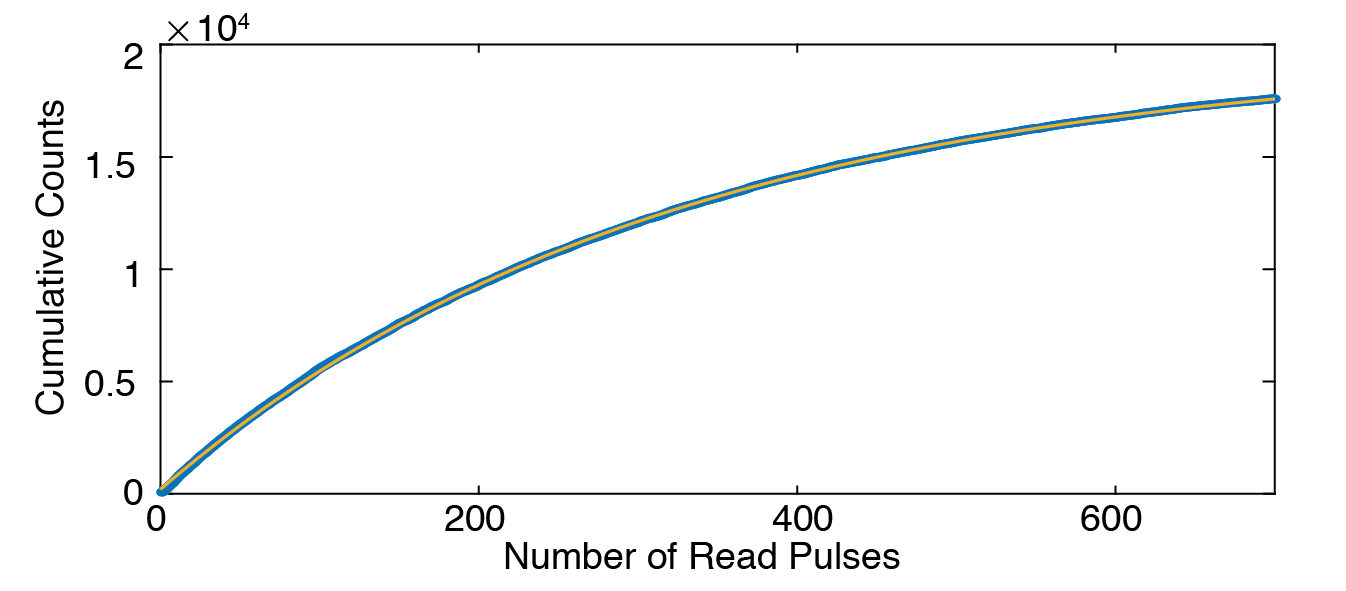}
    \caption[Branching ratio measurement]{Measurement of optical branching ratio extracted from SSRO measurements. The ion is initialized into state $\ket{1}_g$ and the population in $\ket{1}_g$ is measured with a series of readout pulses. The cumulative sum of total counts measured after each read pulse in the experiment is plotted as function of the number of read pulses (blue). Fit (yellow) gives $\beta_{eff} = 0.997$.}
    \label{fig:branchingratiomeasurement}
\end{figure}

The optical branching ratio is measured directly by initializing the ion into $\ket{1}_g$ and measuring the optical pumping of the population as a function of the number of optical read pulses applied. \fig{branchingratiomeasurement} plots the cumulative PL counts, $N_c$, observed as a function of number of read pulses, $N_p$, on $A$ for the ion initialized into $\ket{1}_g$. This is fit to the form
\beq
N_c(N_p) \propto \frac{1-\beta_{eff}^{N_p}}{1-\beta_{eff}},
\eeq
where $\beta_{eff}$ is the effective branching ratio $\beta_{eff} = (1-p_{exc}) + p_{exc}\beta_{parallel}$ to take into account the excitation probability, $p_{exc}$, of the readout pulses. This gives $\beta_{eff} = 0.997$, which is in agreement with the predicted bound on the branching ratio from the lifetime. using $p_{exc}\approx 0.94$ determined from optical Rabi measurements, $\beta_{\parallel} = 0.9968.$ Further improvements to the cyclicity could be achieved with larger Purcell enhancement in cavities with higher quality factors.

\section{Spin initialization}
This section provides additional information on the pulse sequence used to initialize the spin of the single ions (Fig.~1B in main text).

The single ion is first initialized into the qubit subspace by optical pumping out of $\ket{aux}$ on transition $F$, which consists of two $2.5 \,\mathrm{\mu s}$ pulses alternating between the two split transitions discussed earlier (\fig{strainsplitting}) with a total repetition rate of 100 kHz. Transition $F$ is not enhanced by the cavity, but can be driven using light orthogonal to the cavity mode. Once in the excited state $\ket{1}_e$, the ion decays by the cavity-enhanced transition $E$ with high probability to $\ket{0}_g$. The ion is initialized within the qubit subspace by optical pumping on transition $A$, which consists of $2.5 \,\mathrm{\mu s}$ long pulses with a 200 kHz repetition rate. As the optical transition from $\ket{0}_e \rightarrow \ket{0}_g$ is not allowed at zero-field, a microwave pulse is applied simultaneously to the excited state transition $f_e$ during optical pumping on $A$ to create a two-photon transition between $\ket{1}_g$ and $\ket{1}_e$. Once in $\ket{1}_e$, the ion efficiently decays to $\ket{0}_g$ by transition $E$ with branching ratio $\beta_{\parallel}$. This sequence initializes the ion into $\ket{0}_g$. To initialize into $\ket{1}_g$, a microwave $\pi$ pulse is applied on the ground state transition after the optical initialization sequence.

\begin{figure}
    \includegraphics[width=\columnwidth]{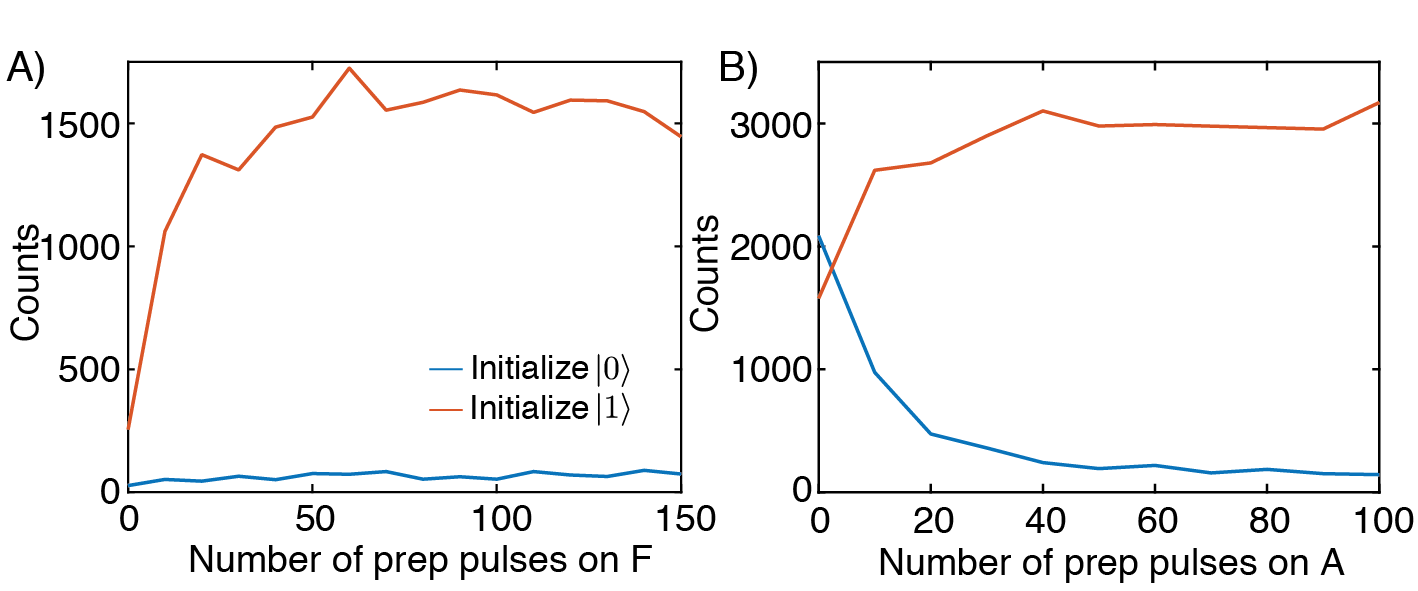}
    \caption[Spin initialization]{Example of spin initialization. Blue (red) scans correspond to preparation into $\ket{0}_g$($\ket{1}_g$).  A) Initialization into qubit subspace. The number of preparation pulses on $A+f_e$ is held fixed at 100 while the number of  preparation pulses on transition $F$ is varied. B) Initialization within qubit subspace. The number of preparation pulses on $F$ is held fixed at 150 while the number of preparation pulses on transition $A+f_e$ is varied.}
    \label{fig:Opticalpumping}
\end{figure}

To demonstrate and assess the quality of the spin initialization scheme, the population in $\ket{1}_g$ is measured for varying lengths of the preparation sequences. \fig{Opticalpumping}A shows optimization of optical pumping out of the $\ket{aux}$ state by varying the number of pulses on $F$, while keeping the number of initialization pulses on $A + f_e$ fixed at 100. From the observed count rate, optical branching ratio and detection efficiency, the initialization into the qubit subspace is estimated to be $>95 \%$. \fig{Opticalpumping}B shows initialization into $\ket{1}_g$ (red) or $\ket{0}_g$ (blue) as the number of pulses on $A + f_e$ is increased while holding the number of pulses on $F$ fixed at 150. Without any subtraction of background count contributions, a population contrast of $91\%$ is observed, which corresponds to an initialization fidelity of $96\%$ within the qubit subspace. This demonstrates that this pumping scheme allows for efficient initialization between these two spin states in under $500 \; \mathrm{\mu s}$.  Here, the state population in $\ket{1}_g$ is measured using PLE with a series of 500 $\pi$ pulses on transition $A$. The initialization measured in this way will be limited by the readout fidelity of this pulse sequence, so represents a lower bound.

\section{Spectral diffusion and post-selection}
\begin{figure}
     \includegraphics[width=\columnwidth]{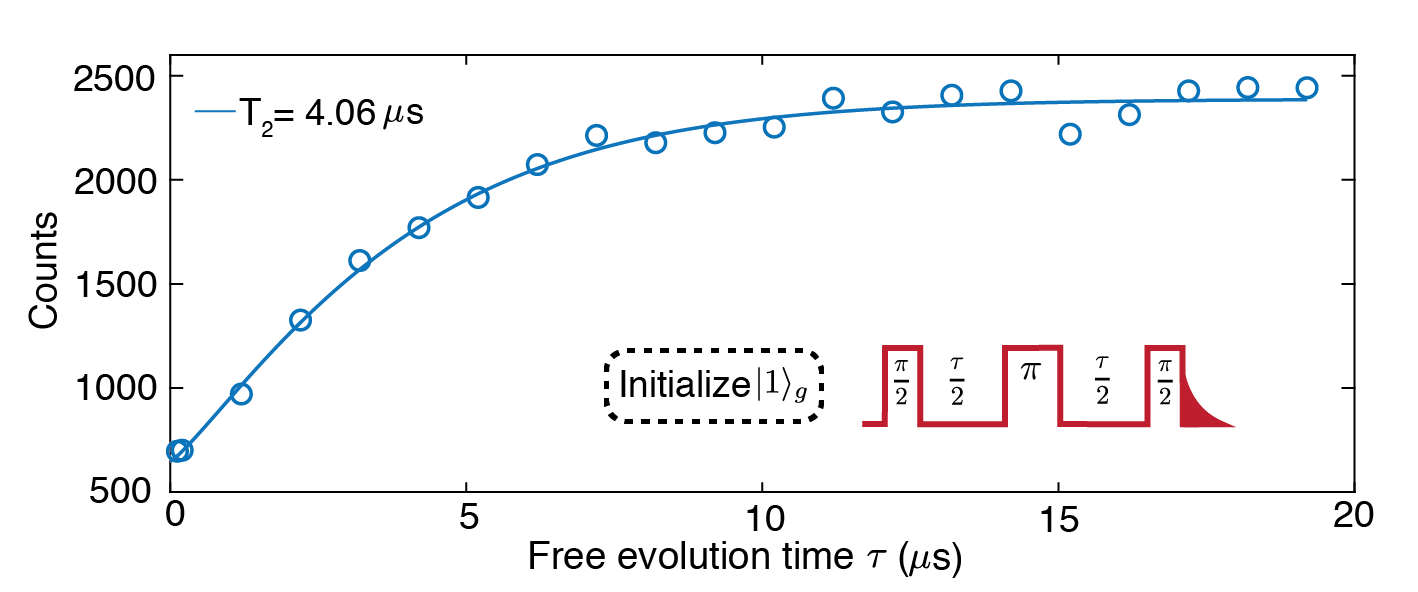}
    \caption[Optical echo measurement]{Measurement of optical $T_2$ on ion Y using an echo sequence (inset). Fit gives $T_2 = 4.06\ \mathrm{\mu s}$. }
    \label{fig:opticalecho}
\end{figure}

\fig{opticalecho} shows an optical echo measurement on transition $A$, which gives a coherence time of $T_{2,o} = 4.06\ \mathrm{\mu s}$. The considerably shorter optical Ramsey coherence time (Fig.~2E) indicates that the measured $T_{2,o}^*$ is limited by quasi-static fluctuations in the frequency of transition $A$. The current $T_{2,o}^*$ will be detrimental to photon indistinguishability, but can be improved, for instance, by using post-selection to ensure the ion is on resonance with the excitation pulse \cite{Robledo2010}. We demonstrate the possibility of this approach in this system by post-selecting Ramsey measurements based on number of photons, $n_{c}$ detected during a subsequent probe sequence consisting of a series of resonant, low power optical pulses on the transition $A$. 

\begin{figure}
     \includegraphics[width=\columnwidth]{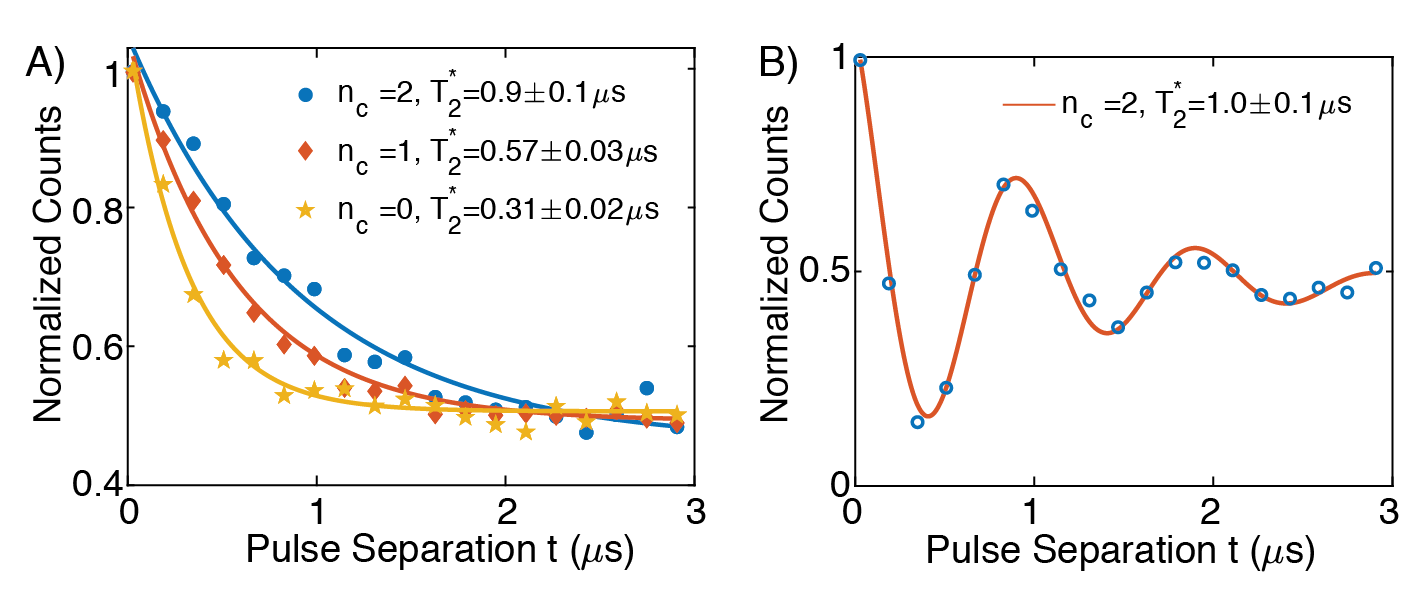}
    \caption[Post-selected optical Ramsey]{A) Post-selected optical Ramsey decays showing improvement in $T_2^*$ for increasing numbers of photons $n_{c}$ detected in a subsequent probe sequence. B) Post-selected Ramsey sequence for $n_{c} = 2$ with readout detuned by 1 MHz to demonstrate that decay is due to optical coherence.}
    \label{fig:opticalpostselection}
\end{figure}

\fig{opticalpostselection}A shows the results of postselected resonant Ramsey measurements, where improvements in $T_{2,o}^*$ are observed for increasing number of probe photons detected. \fig{opticalpostselection}B shows similar measurements with the excitation pulses detuned by 1 MHz to give rise to characteristic Ramsey fringes, verifying that this indeed corresponds to a coherence decay. Post-selecting with $n_c=2$ leads to a $T_{2,o}^*$ of $1.0 \pm 0.1 \ \mu s$ albeit with approximately $84\%$ of the Ramsey experiments discarded. Higher collection efficiency in future devices should enable further improvements in $T_{2,o}^*$ using this or similar post-selection technique. 

One possible cause of these quasi-static fluctuations in the optical transition frequency is the magnetic dipole-dipole, or superhyperfine (SHF), interaction between the Yb electron spin and host nuclei, specifically vanadium ($\mathrm{I_V}=7/2$) and yttrium ($\mathrm{I_Y}=1/2$). Coupling to the two nearest vanadium ions is expected to dominate due to the $1/r^3$ scaling of the magnetic dipole-dipole interaction and the relative size of the nuclear g-factors ($\mathrm{g_V}=1.5$ and $\mathrm{g_Y}=-0.27$) \cite{Car2018}. Simulations of the optical transition that take into account the superhyperfine interaction give a broadening on $A$ of < 50 kHz (FWHM). This does not fully account for the observed 370 ns $T_{2,o}^*$ time, which corresponds to an 860 kHz linewidth. Further investigation of the limits of the $T_{2,o}^*$ due to the SHF mechanism and other factors, such as the second-order DC stark shift of this transition, will be the subject of future research.

\section{Additional spin measurements}

\subsection{Optically-detected magnetic resonance}
\begin{figure}
     \includegraphics[width=\columnwidth]{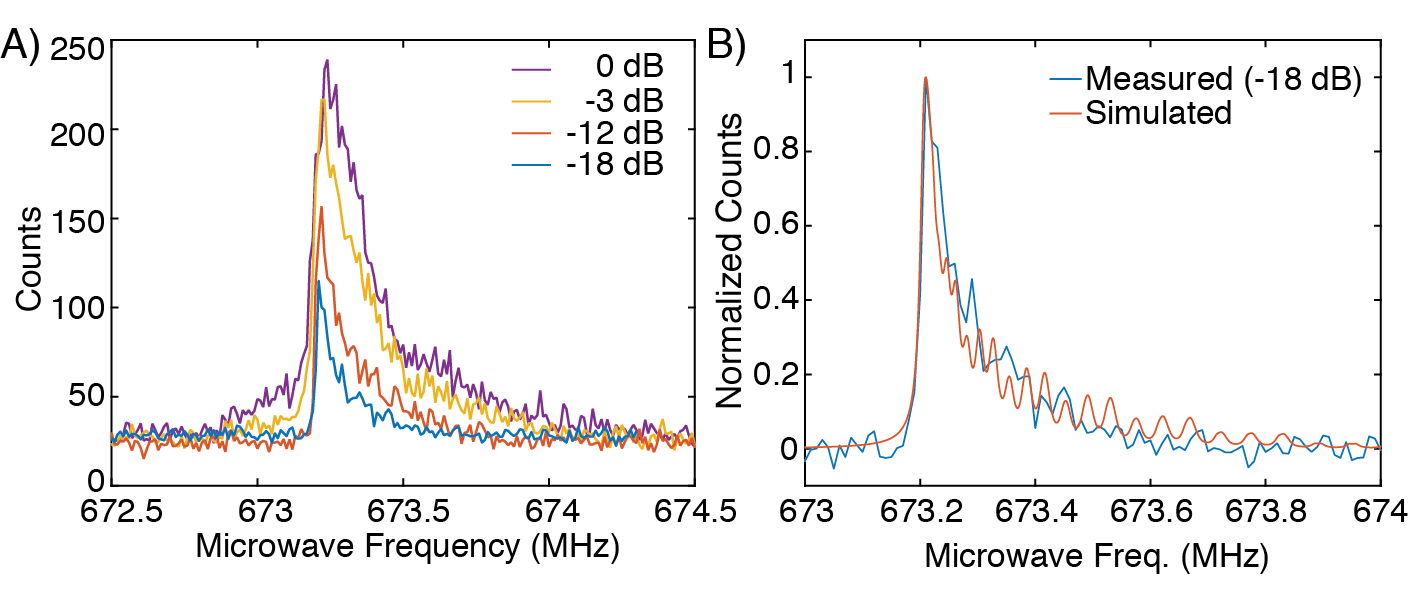}
\caption[Ground-state ODMR]{A) Optically-detected magnetic resonance on the $\ket{0}_g \rightarrow \ket{1}_g$ transition. The ion is optically pumped into $\ket{0}_g$, a microwave pulse is applied, and the population in $\ket{1}_g$ is optically read out as a function of microwave frequency. The ODMR signal is measured with increasing attenuation of the microwave pulse to minimize the effects of power broadening. B) Simulated ODMR spectrum (red) that takes into account superhyperfine interaction with 4 neighbors (3V, 1Y) to verify the form of the experimentally observed ODMR signal (blue) is expected. Here, each individual simulated transition is plotted with 15 kHz of broadening.}
\label{fig:171s_gndODMR}
\end{figure}

Optically detected magnetic resonance (ODMR) measurements were performed on the ground state spin transition $\ket{0} \leftrightarrow \ket{1}$ for initial calibration of the spin transition frequencies and to bound the coherence time. For this measurement, the ion is initialized into state $\ket{0}$, a $160 \ \mu s$ long microwave pulse is applied, and the population in state $\ket{1}_g$ is read out optically. \fig{171s_gndODMR} shows a series of ODMR frequency scans on ion Y performed at successively lower microwave powers to reduce the effects of power broadening. At the lowest microwave power used, we measure a linewidth (full-width-half-maximum) of 48 kHz. This places a lower bound on the spin $T_{2,s}^*$ time of $6.6 \ \mu s$, which in agreement with the spin $T_{2,s}^*$ measured directly using a Ramsey sequence. From this and similar measurements on ion X, we extract the qubit transition frequencies of ion X and Y to be 674.48 MHz and 673.24 MHz respectively (the inhomogeneous linewidth of this transition measured in 100 ppm \ybyvo is <1 MHz).

The asymmetric profile of the ODMR spectrum is attributed to second-order perturbations of the qubit transition by magnetic dipole-dipole (superhyperfine) interactions with nearby nuclear spins (yttrium and vanadium). To verify this this asymmetry is due to the superhyperfine interaction, the energy level structure of the $\ket{0}_g \rightarrow \ket{1}_g$ transition was modeled by introducing magnetic dipole-dipole coupling of the Yb electron spin to neighboring nuclear spins \cite{Car2018}. \fig{171s_gndODMR} plots the simulated spectrum due to coupling with 3 nearest vanadium and 1 nearest yttrium, which shows good agreement with the experimental results. 

\subsection{Calibration of pulses and readout}

Additional calibration of the center frequency of the spin transition is accomplished by minimizing the frequency of Rabi oscillations as a function of microwave drive frequency. 

The length of microwave control pulses is extracted from the Rabi oscillations. Finer calibrations of $\pi$ pulse lengths are performed by initializing the ion into $\ket{0}_g$, applying an even number of $\pi$ pulses, and minimizing the resulting population in $\ket{1}_g$ as a function of pulse length.

For the spin coherence measurements presented in the text, the phase of the final $\pi/2$ pulse is chosen to be $180^\circ$ out of phase with the initial $\pi/2$ pulse to map the coherence to a population on the $\ket{0}_g$. This gives rise to the increasing exponential decay observed in Fig.~3. 

Unless otherwise specified, optical readout of the spin-state for coherence and lifetime measurements is performed by measuring the fluorescence observed in a single series of readout pulses.

\subsection{Magnetic field dependence of spin coherence}
As described inthe main text, at zero field the hyperfine interaction gives rise to mixed electron-nuclear spin states of the form $\ket{1} = \frac{1}{\sqrt{2}} \left(\ket{ \uparrow \Downarrow } + \ket{ \downarrow \Uparrow } \right)$ and  $\ket{0} = \frac{1}{\sqrt{2}} \left(\ket{ \uparrow \Downarrow } - \ket{ \downarrow \Uparrow } \right)$. These states have zero net magnetic moment and are thus first-order insensitive to perturbations by the Zeeman interaction. Such a transition is often referred to in the rare-earth literature as a ZEFOZ transition (as it has zero first-order Zeeman shift) \cite{McAuslan2012}. This reduces the sensitivity of transitions involving these states to magnetic field fluctuations that arise from host nuclei and other rare-earth ions in the crystal \cite{Ortu2017,Rakonjac2018a}. In this case, the magnetic field sensitivity of these transitions, and thus, coherence time is determined by the second-order Zeeman interaction \cite{McAuslan2012}.

\begin{figure}[h]
    \centering
    \includegraphics{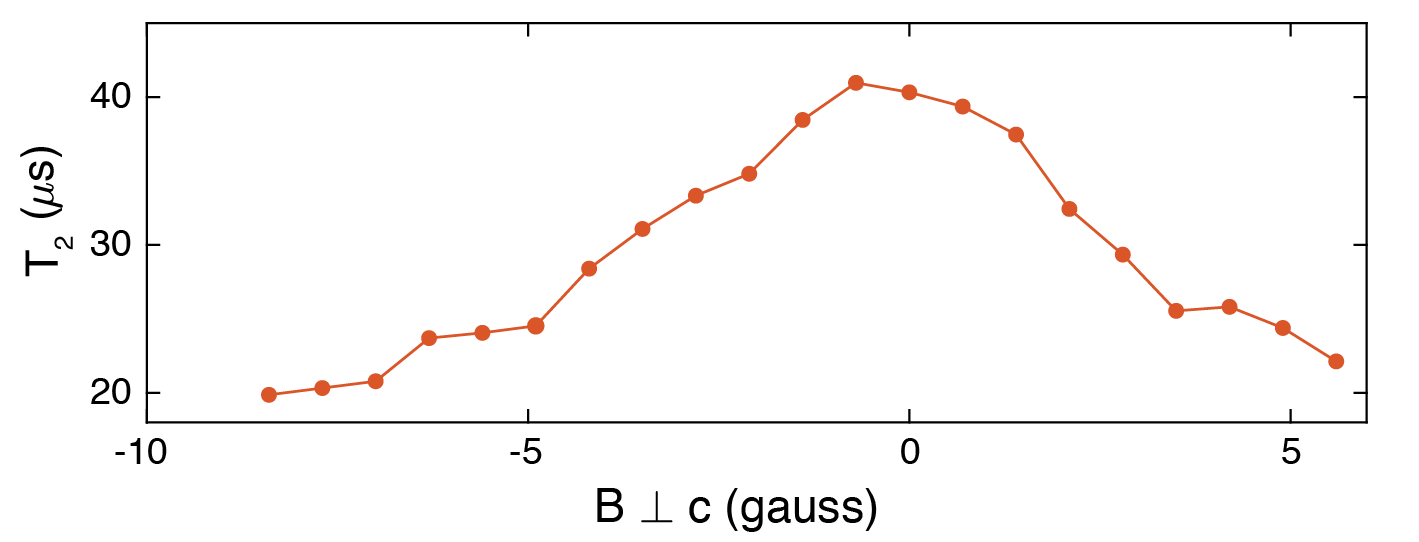}
    \caption[Magnetic field dependence of spin coherence]{Coherence of $\ket{0}_g \leftrightarrow \ket{1}_g$ measured using a spin-echo sequence for varying magnetic fields applied perpendicular to the $c$-axis of the crystal.}
    \label{fig:coherencetimewithfield}
\end{figure}

As an external DC magnetic field is applied, the spin Hamiltonian predicts an increased magnetic field sensitivity for these transitions and thus, a reduction in coherence time \cite{Ortu2017}. This is investigated in \fig{coherencetimewithfield}, which shows the measured qubit coherence time for a magnetic field applied along the $a$-axis. A decrease in coherence time is observed away from zero field, which provides confirmation that magnetic field sensitivity plays a dominant role in the coherence time.

\subsection{Noise spectroscopy and dynamical decoupling}
\begin{figure}[hb]
    \includegraphics[width=1\columnwidth]{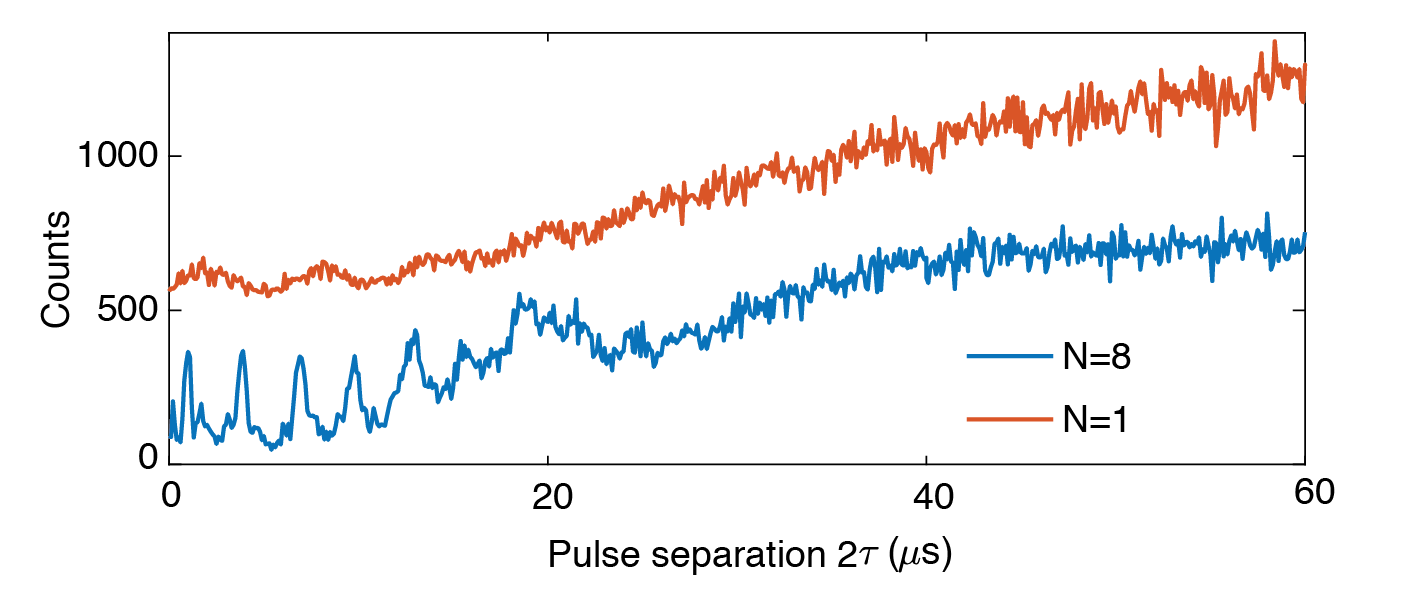}
    \caption[Fine CPMG scans]{Fine resolution CPMG scans performed with $N=1$ (red) and $N=8$ (blue) rephasing pulses showing periodic collapses and revivals of the spin coherence. The $N=1$ scan is offset by 500 counts for clarity.}
    \label{fig:cpmgnoisespectroscopy}
\end{figure}

This section describes addition measurements performed to understand and prolong the spin coherence. The red trace in \fig{cpmgnoisespectroscopy} shows a Hahn echo measurement on the $\ket{0}_g$ – $\ket{1}_g$ transition with a coherence time of $T_{2,s} = 43.5 \, \mathrm{\mu s}$. The stretched-exponential decay is indicative of slow fluctuations in the ion’s magnetic environment \cite{Klauder1962}, which is attributed to a nuclear spin bath consisting of lattice spins (specifically $\mathrm{I_V}=7/2$ and $\mathrm{I_Y}=1/2$). The coherence time in this case can be extended by applying dynamical decoupling (DD) pulse sequences such as the Carr-Purcell-Meiboom-Gill (CPMG) pulse sequence, which takes the form \cite{Meiboom1958a}
 \beq
 \of{\frac{\pi}{2}}_y - \of{\tau - \pi_x  - \tau}^N -  \of{\frac{\pi}{2}}_{-y}.
 \eeq
This sequence can be thought of as a multi-pulse analog to the spin echo technique that extends coherence times by high-pass filtering environmental noise on a frequency scale $\omega \sim 1/\tau$. This sequence can also be treated as a frequency-domain noise filter of the form derived in \cite{Cywinski2008}.

The blue trace in \fig{cpmgnoisespectroscopy} shows a measurement of spin coherence using a CPMG pulse sequence with eight $\pi$ pulses, plotted as a function of the $\pi$ pulse separation time $2\tau$. Periodic collapses and revivals in CPMG coherence are observed, which is indicative of a narrow-band noise source at 340 kHz. This is currently attributed to coupling to nearby nuclear spins (similar signatures have been observed with nitrogen vacancy centers weakly coupled to $^{13}$C nuclear spins \cite{Zhao2012,Taminiau2012, Kolkowitz2012}), but further investigations will be the subject of future work. By operating with pulse separations at integer multiples of the coherence revival time, the spin-qubit is effectively decoupled from this narrowband noise as shown in Fig.~3C.

\begin{figure}[hb]
    \includegraphics[width=\columnwidth]{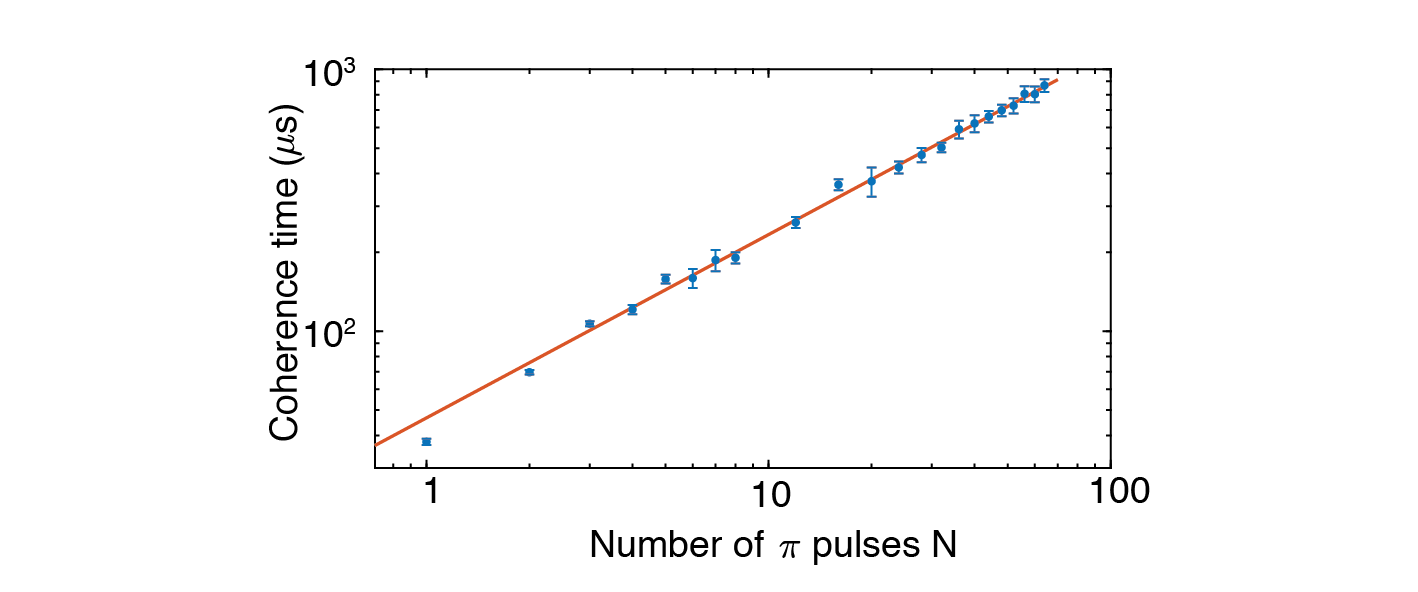}
    \caption[Scaling of CPMG coherence ]{Scaling of coherence time extracted from CPMG envelope (Fig.~3) for varying numbers of rephasing pulses. Fit gives $T_{2,s}^N \propto N^{0.70\pm0.01}$.}
    \label{fig:cpmgscaling}
\end{figure}

To further study the coherence time limitations of this transition, we follow the procedure outlined in \cite{Medford2012}. In \fig{cpmgscaling}, the CPMG coherence time is plotted as a function of the number of $\pi$ pulses in the CPMG sequence ($N$). We extract a power-law dependence of the form $T_{2,s} \propto N^{0.70\pm0.01}$. This indicates a noise spectral density of the form $S(\omega)\propto\omega^{-2.3\pm0.1}$, which agrees well with the expected $ S(\omega)\propto\omega^{-2}$ for coupling to a dipolar spin bath approximated by a classical source of Ornstein-Uhlenbeck noise \cite{Dobrovitski2009}.

The combined requirements of decoupling from narrowband noise and minimizing the pulse separation time to filter out $1/\omega^2$ noise identifies an optimal pulse separation of $2\tau=5.74$ $\mathrm{ \mu s}$. By measuring the decay in coherence as a function of CPMG pulse number at this optimal pulse separation, we obtain a $T_{2,s}$ time of 30 ms (Fig.~3D).

\begin{figure}[hb]
    \centering
    \includegraphics[width=\columnwidth]{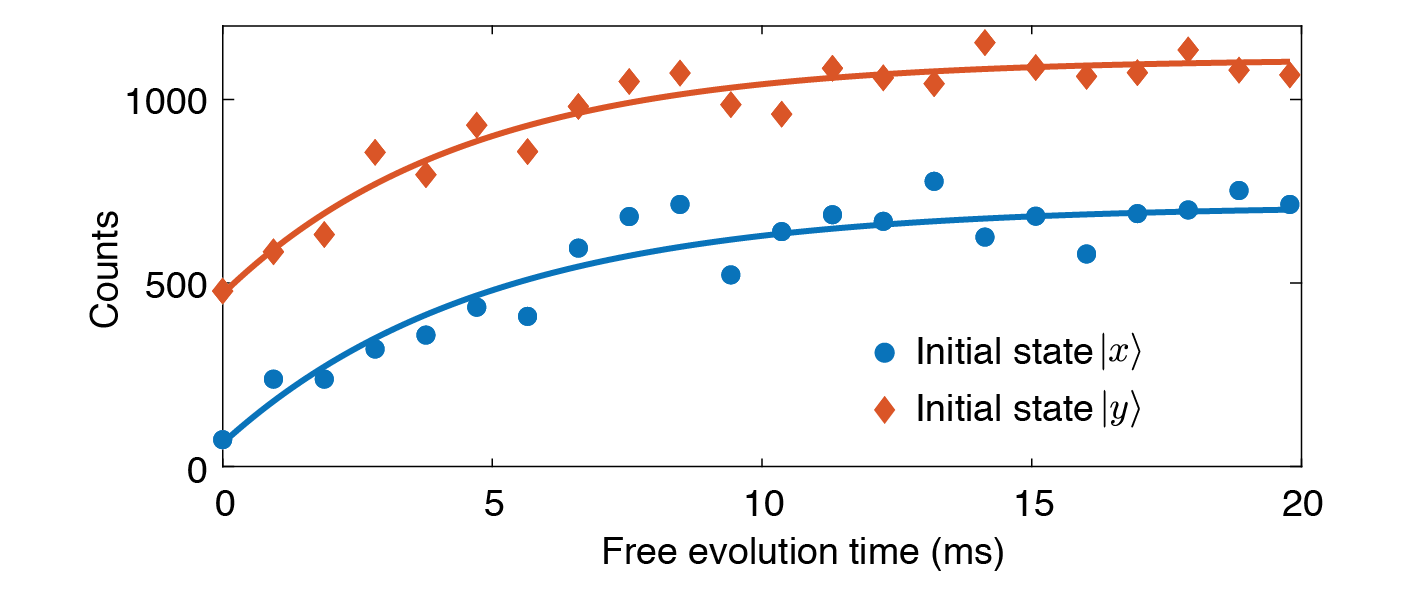}
    \caption[Coherence decay using XY-8 sequence.]{Decay of coherence using XY-8 sequence for two phases of the initial superposition state. Scans are offset by 500 counts for clarity.}
    \label{fig:xy8measurements}
\end{figure}

The CPMG sequence performs single-axis decoupling; it is only robust to pulse errors for initial states parallel to the x-axis of the Bloch sphere. CPMG sequences are thus unsuitable for preservation of arbitrary quantum states. For instance, repeating this sequence with an initial $(\pi/2)_x$ pulse (i.e. y-directed initial state) leads to a reduction in coherence time by two orders of magnitude due rapid accumulation of pulse errors ($T_{2,s}=240\ \mathrm{\mu s}$). We can mitigate this effect by using sequences with decoupling pulses applied along multiple axes \cite{Gullion1990}. As an example of this, \fig{xy8measurements} shows measurements of coherence times up to 4 ms using the XY-8 pulse sequence:
\beq
\left(\frac{\pi}{2}\right)-[\tau\mhyphen\pi_x\mhyphen2\tau\mhyphen\pi_y\mhyphen2\tau\mhyphen\pi_x\mhyphen2\tau\mhyphen\pi_y \mhyphen 2\tau\mhyphen \pi_y\mhyphen 2\tau\mhyphen\pi_x\mhyphen2\tau\mhyphen\pi_y\mhyphen2\tau\mhyphen\pi_x\mhyphen\tau]^N-\left(\frac{\pi}{2}\right).
\eeq
This sequence has been demonstrated to preserve coherence for arbitrary initial states \cite{deLange2010}, which is a crucial requirement for using this transition as a quantum memory.

\subsection{Spin lifetime and device temperature}
\label{sec:lifetimeandtemp}

\begin{figure}
     \includegraphics[width=\columnwidth]{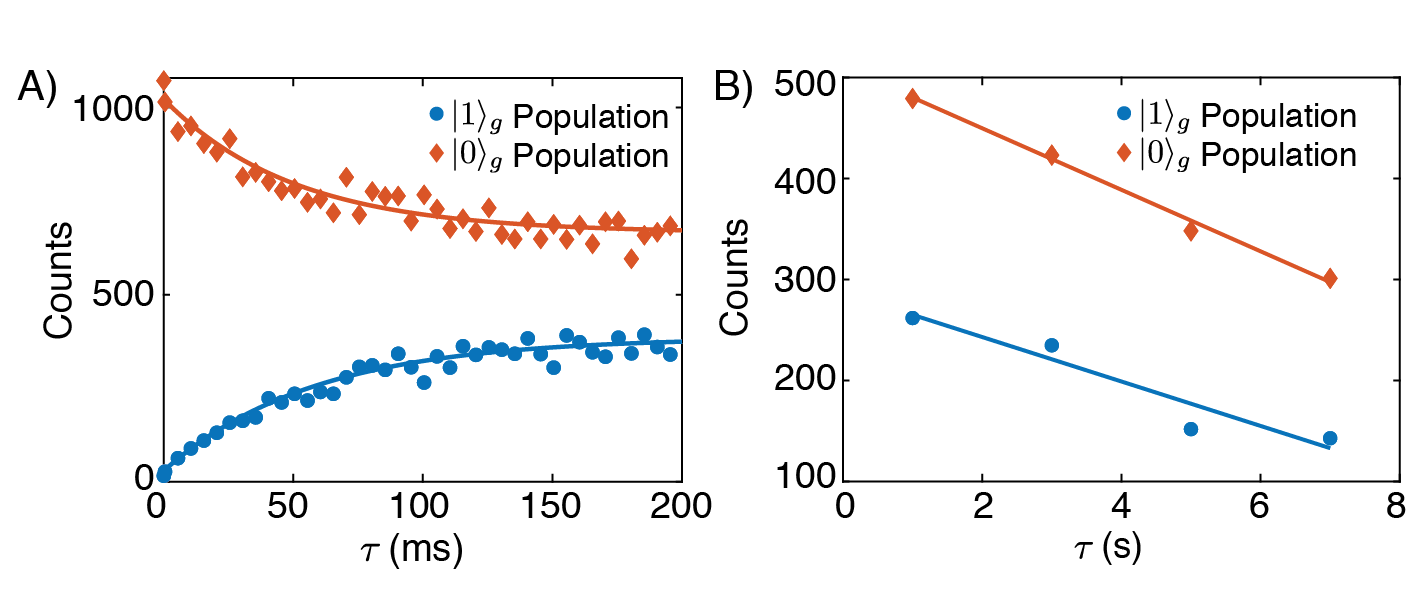}
    \caption[Spin lifetime]{A) Measurement of population in $\ket{0}_g$ and $\ket{1}_g$ after initializing into $\ket{0}_g$ and waiting for time $\tau$. Exponential fit gives lifetime of $54\pm 5 \, \mathrm{ ms}$. The measured population difference corresponds to a device temperature of $59\pm 4\,\mathrm{ mK}$. B) Slow decay of population from qubit subspace into $\ket{aux}_g$ with decay constant $26\pm 3 \, \mathrm{ s}$.}
    \label{fig:spinlifetime}
\end{figure}

The lifetimes of the spin transitions are measured by initializing the ion into state $\ket{0}_g$, waiting for a period of time $\tau$ and optically reading out either the $\ket{0}_g$ or $\ket{1}_g$ population on transition $A$. The $\ket{0}_g$ population is measured by applying a microwave $\pi$ pulse followed by optical readout on transition $A$. \fig{spinlifetime} shows the results of such a measurement at a cryostat temperature of 40 mK. A bi-exponential decay of the $\ket{0}_g$ population is observed: a rapid decay ($54\pm 5\,\mathrm{ms}$) due to thermalization with state $\ket{1}_g$ (\fig{spinlifetime}A) followed by a much slower decay ($26\pm 3\, \mathrm{s}$) as the states $\ket{0}_g$ and $\ket{1}_g$ thermalize with $\ket{aux}_g$ (\fig{spinlifetime}B). From the ratio between the $\ket{0}_g$ and $\ket{1}_g$ populations after the short $\ket{0}_g \leftrightarrow \ket{1}_g$ relaxation, we extract a device temperature of $59\pm 4\,\mathrm{mK}$. This agrees well with the measured mixing chamber plate temperature of 40 mK.

Similar measurements were performed at cryostat temperatures up to 1.2 K. We observe less than a factor of two change in the decay rate on the $\ket{0}_g \leftrightarrow \ket{1}_g$ transition over this range, indicating that this is unlikely to be dominated by a phonon-assisted process that would be expected to scale strongly with temperature \cite{Cruzeiro2017}. Instead, we postulate a direct spin-spin relaxation mechanism mediated by magnetic dipole-dipole interactions with other $^{171}\mathrm{Yb}^{3+}$ ions in the crystal. This is expected to be a dominant effect in this system due to the narrow inhomogeneous linewidths of the spin transition at zero field (<1 MHz). Furthermore, the large difference between $\ket{0}_g  \leftrightarrow \ket{1}_g$ and $\ket{0}_g, \ket{1}_g   \leftrightarrow \ket{aux}_g$ relaxation rates is in agreement with the $g^4$ scaling for this mechanism \cite{Bottger2006} (corresponding $g$-factors for these two transitions are -6.08 and 0.85, respectively \cite{Ranon1968}).

More measurements are necessary to further investigate the underlying relaxation mechanism. For lifetimes dominated by direct spin flip-flops, we expect that reducing the population of resonant Yb-171 ions in the $\ket{0}_g, \ket{1}_g$ manifold will increase the $\ket{0}_g  \leftrightarrow \ket{1}_g$ lifetime thereby also improving the CPMG coherence time that is currently approaching the lifetime limit. This could be accomplished by optically initializing other Yb-171 ions in the device into the $\ket{aux}_g$ state, or investigating samples with lower doping concentrations and increased spin inhomogeneity \cite{Welinski2017_sc}.

\section{Single-shot readout fidelities}


\subsection{Photon count distributions and readout fidelity}
For single-shot readout, the state of the ion is assigned based on the number of photons detected during an optical readout period consisting of a series of optical excitation pulses on transition $A$.  We choose a photon number cutoff, $n_c$, and assign the state of the ion to $\ket{1}$ if we measure $\ge n_c$ photons during the readout period and to $\ket{0}$ is we measure $<n_c$ photons. As mentioned earlier, the qubit is optically read out on transition $A$, because transition $E$ overlaps with the optical transition of the zero-spin isotope.

The fidelity $F_i$ of this readout is the probability of obtaining the measurement outcome $i$ after initializing into $i$: $F_i = 1-\epsilon_i$, where $\epsilon$ is the error of assignment given by: 
\beq
\epsilon_0 = \sum_{k = n_c}^\infty P_{\ket{0}}(k) 
\eeq
 and 
\beq
\epsilon_1 = \sum_{k = 0}^{n_c - 1} P_{\ket{1}}(k).
\eeq
Here, $P_{\ket{i}}(k)$ is the photon count distribution describing the probability of measuring $k$ counts with the ion in $\ket{i}$.

The photon count distribution for the ion initialized in $\ket{0}_g$ will be determined by the background count rate $\Gamma_{bg}$ due to detector dark counts, light leakage, or fluorescence from other ions in the crystal. We assume these background counts follow a Poisson distribution with the average photon number $\bar{N}_{bg} = \Gamma_{bg} N_{r} t_{r}$ detected in $N_{r}$ readout pulses with integration time $t_{r}$ per pulse. 

The photon count distribution for the ion in $\ket{1}_g$ will be a convolution of the counts due the ion and the background:
\beq
P_{\ket{1}_g}(N_{tot} = n) = \sum_{k = 0}^n P(N_{ion} = k) P(N_{bg} = n-k),
\eeq
Where $N_{tot} = N_{bg} + N_{ion}$ is the total number of counts measured due to counts from the ion, $N_{ion}$, and the background $N_{bg}$. The photon count distribution from the ion will be determined by the cyclicity of transition $A$ and overall detection efficiency. Upon excitation, the ion has a probability of optically decaying to $\ket{aux}$ given by $p_f = 1-\beta_{eff}$, where $\beta_{eff}$ is the effective branching ratio for decay on the $\ket{0}_e \rightarrow \ket{1}_g$ transition discussed in \sect{branchingratio}. The total number of successful read pulses $N_r$ during a given readout sequence before the ion is optically pumped follows a geometric distribution: 
\beq
P(N_r = k) = (1-p_f)^{k} p_f,
\label{eq:geometricdist}
\eeq
where $k = \{0, 1, 2, 3, ...\}$. 
As a result, the total number of counts due to the ion, $N_{ion}$, will be a randomly stopped sum of $N_r$ pulses that are sampled binomially with total system detection efficiency $p_{tot}$:

\begin{align}
    P(N_{ion}= k) &= \sum_{N_r = k}^\infty P (N_{ion} = k | N_r) P(N_r ) \\
    &= \sum_{N_r = k}^\infty \binom{N_r}{k} p_{tot}^k (1-p_{tot})^{(N_r - k)} (1-p_f)^{N_r } p_f .
\end{align}
This can be written as another geometric distribution 
\beq
P(N_{ion}= k) = (1-p_n)^{k} p_n,
\eeq
where $p_n = \frac{p_f}{p_{tot} + p_f -p_{tot} p_f}$.

The form above assumes that readout is done using a sufficient number of read pulses such that the ion is optically pumped away by the end of the sequence with high probability and we can thus approximate the distribution as geometric. To account for a finite number of read pulses, $N_{max}$, \eq{geometricdist} can be replaced with a truncated geometric distribution: 
\begin{align}
P(N_r= k | N_{max}) &= \frac{P(N_{r} = k)}{\sum_{j = 0}^{N_{max}}P(N_{r} = j)} \\
&= \frac{(1-p_f)^{k} p_f}{1-(1-p_f)^{N_{max}+1}}.
\end{align}

To measure the photon count distribution and assign a readout fidelity, the ion is initialized into $\ket{0}_g$ or $\ket{1}_g$ and the readout procedure is repeated many time to acquire adequate statistics. Fig.~4b shows one such photon count histogram where 400 read pulses were used per sequence, which is in good agreement with the expected form for the distributions given above. Using a cutoff of $n_c=1$, we extract $F_0 = 96.1\%$ and $F_1 = 64.0\%$ for an average fidelity of $F_{avg} = (F_0 + F_1)/2 = 80.0\%$. This fidelity is not strictly the readout fidelity, but the combined initialization and readout fidelity. The low dark and leakage counts in this system enables high-fidelity readout when the ion is initialized into $\ket{0}_g$, but the geometric distribution of ion counts due to the finite branching ratio and detection efficiency limits $F_1$. Furthermore, detection of zero photons during the readout period does not distinguish between the ion being in $\ket{aux}$ or $\ket{0}$. 

\subsection{Conditional readout fidelities}
We implement a conditional readout procedure described in the text to overcome the difference in readout fidelity between the qubit states and improve the overall readout fidelity. We read out the state of the ion using a series of readout pulses, apply a $\pi$ pulse to the $\ket{0}_g \rightarrow \ket{1}_g$ transition, and readout the state of the ion again. The state of the ion is determined during each readout period as described above and labeled as $\ket{a b}$, where $a$($b$) is the outcome of the first (second) readout. The initial state is conditionally assigned to $\ket{0}_g$ on the observation of $\ket{01}$ and to $\ket{1}_g$ on the observation of $\ket{10}$. This procedure takes advantage of the high $F_0$ fidelity, which corresponds to a high probability of the ion being in $\ket{1}_g$ if one or more photons is detected. Conditionally assigning the state on at least one count in the first or second readout period also ensures that the ion was in the qubit subspace during the measurement.

The expected fidelity of the conditional readout can be determined by the fidelities $F_i$ of the first and second readout sequence (Table \ref{table:dualSSROreadoutfidelities}). These forms assume that the readout fidelity is identical for the first and second readout sequences, but in general the overall fidelity of the first readout will be higher due to spin relaxation and microwave pulse errors. The conditional readout fidelity is
\beq
F_{cond} = \frac{F_{0} F_{1}}{F_{0} F_{1} + (1 - F_{1})(1-F_{0})}.
\eeq
The improved fidelity comes at reduced success rate of detecting the ion in either qubit state:
\beq
P_{success} = F_{0} F_{1} + (1 - F_{1})(1-F_{0}).
\eeq

\begin{table}[h]
\centering
\caption[Conditional readout probabilities]{Probability of outcome $\ket{ab}$ of conditional single-shot readout protocol for ion prepared in $\ket{i}$ in terms of readout fidelities for a single series of pulses. These forms assume identical readout fidelities for the first and second readout sequences.}
\begin{tabular}{|c||c|c|}
\multicolumn{1}{l}{Measured} & \multicolumn{2}{c}{Prepared }   \\
\hline
$\ket{ab}$  & $\ket{0}$  & $\ket{1}$  \\ \hline
$\ket{0 0}$ & $F_{0}(1-F_{1})$      & $(1-F_{1}) F_{0}$  \\
$\ket{0 1}$ & $F_{0} F_{1}$      & $(1 - F_{1})(1-F_{0})$     \\
$\ket{1 0}$ & $(1 - F_{0})(1-F_{1})$ & $F_{1} F_{0}$      \\
$\ket{1 1}$ & $(1 - F_{0})F_{1}$ & $F_{1} (1 - F_{0})$   \\
\hline
\end{tabular}
\label{table:dualSSROreadoutfidelities}
\end{table}

\begin{figure}
    \includegraphics[width=1\columnwidth]{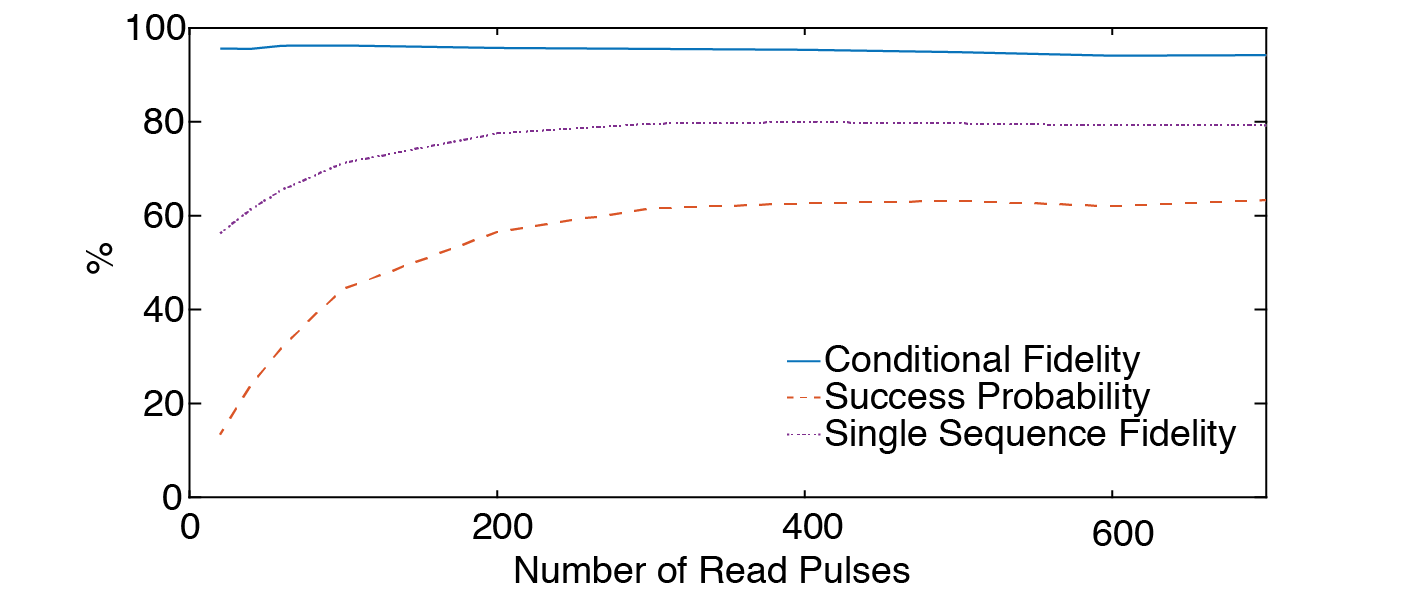}
    \caption[Single-shot readout fidelity]{Dependence of single-shot readout fidelity as a function of number of read pulses for the single readout sequence (purple, dot-dashed) and the conditional sequence (blue, solid). The success probability of observing the ion within the qubit subspace (i.e. measuring $\ket{10}$ or $\ket{01}$) using the conditional readout procedure is shown in red (dashed).}
    \label{fig:SSROwithpulsenumber}
\end{figure}

\fig{SSROwithpulsenumber} shows the fidelity of the single and conditional readout along with the success rate for increasing numbers of readout pulses. For the data presented in Figs.~4B and 4C, 400 read pulses are used in each sequence, which corresponds to the plateau in the success rate with $F_{1,avg} = 80\%$, $F_{cond} = 95.3 \%$, and $P_{success} = 62.6\%$. Slightly higher fidelities are achieved with fewer pulses at a reduced success rate. Further improvements to the readout fidelity will be enabled by higher overall detection efficiencies and by further enhancements of the branching ratio, which should be achievable in a higher Q cavity.

\bibliographystyle{Science}
\bibliography{library}

\begin{thebibliography}{10}

\bibitem{TZhong2016}
T.~Zhong, J.~Rochman, J.~M. Kindem, E.~Miyazono, A.~Faraon, {\it Optics
  Express\/} {\bf 24}, 536 (2016).

\bibitem{Tiecke2014}
T.~G. Tiecke, {\it et~al.\/}, {\it Nature\/} {\bf 508}, 241 (2014).

\bibitem{Marsili2013}
F.~Marsili, {\it et~al.\/}, {\it Nature Photonics\/} {\bf 7}, 210 (2013).

\bibitem{Mosor2005}
S.~Mosor, {\it et~al.\/}, {\it Applied Physics Letter\/} {\bf 87}, 10 (2005).

\bibitem{Abragam1970}
A.~Abragam, B.~Bleaney, {\it {Electron Paramagnetic Resonance of Transition
  Ions}\/} (Oxford University Press, Oxford, 2012).

\bibitem{Kindem2018}
J.~M. Kindem, {\it et~al.\/}, {\it Phys. Rev. B\/} {\bf 80}, 1 (2018).

\bibitem{Ortu2017}
A.~Ortu, {\it et~al.\/}, {\it Nature Materials\/} {\bf 17}, 671 (2018).

\bibitem{Macfarlane2007}
R.~M. Macfarlane, {\it Journal of Luminescence\/} {\bf 125}, 156 (2007).

\bibitem{Kitson1998}
S.~C. Kitson, P.~Jonsson, J.~G. Rarity, P.~R. Tapster, {\it Physical Review
  A\/} {\bf 58}, 620 (1998).

\bibitem{Besombes2010}
G.~Sallen, {\it et~al.\/}, {\it Nature Photonics\/} {\bf 4}, 696 (2010).

\bibitem{Schuurmans2000}
F.~J.~P. Schuurmans, P.~D. Vries, A.~Lagendijk, {\it Physics Letters A\/} pp.
  472--477 (2000).

\bibitem{Robledo2010}
L.~Robledo, H.~Bernien, I.~V. Weperen, R.~Hanson, {\it Physical Review
  Letters\/} {\bf 177403}, 1 (2010).

\bibitem{Car2018}
B.~Car, L.~Veissier, A.~Louchet-Chauvet, J.~L. {Le Gou{\"{e}}t},
  T.~Chaneli{\`{e}}re, {\it Physical Review Letters\/} {\bf 120}, 1 (2018).

\bibitem{McAuslan2012}
D.~L. McAuslan, J.~G. Bartholomew, M.~J. Sellars, J.~J. Longdell, {\it Physical
  Review A\/} {\bf 85}, 032339 (2012).

\bibitem{Rakonjac2018a}
J.~V. Rakonjac, Y.-H. Chen, S.~P. Horvath, J.~J. Longdell, {\it
  http://arxiv.org/abs/1802.03862\/}  (2018).

\bibitem{Klauder1962}
J.~R. Klauder, P.~Anderson, {\it Physical Review\/} {\bf 125} (1961).

\bibitem{Meiboom1958a}
S.~Meiboom, D.~Gill, {\it Review of Scientific Instruments\/} {\bf 688} (1958).

\bibitem{Cywinski2008}
{\L}.~Cywi{\'{n}}ski, R.~M. Lutchyn, C.~P. Nave, S.~{Das Sarma}, {\it Physical
  Review B\/} {\bf 77}, 1 (2008).

\bibitem{Zhao2012}
N.~Zhao, {\it et~al.\/}, {\it Nature Nanotechnology\/} {\bf 7}, 657 (2012).

\bibitem{Taminiau2012}
T.~H. Taminiau, {\it et~al.\/}, {\it Physical Review Letters\/} {\bf 109},
  137602 (2012).

\bibitem{Kolkowitz2012}
S.~Kolkowitz, Q.~P. Unterreithmeier, S.~D. Bennett, M.~D. Lukin, {\it Physical
  Review Letters\/} {\bf 109}, 1 (2012).

\bibitem{Medford2012}
J.~Medford, {\it et~al.\/}, {\it Physical Review Letters\/} {\bf 108}, 086802
  (2012).

\bibitem{Dobrovitski2009}
V.~V. Dobrovitski, A.~E. Feiguin, R.~Hanson, D.~D. Awschalom, {\it Physical
  Review Letters\/} {\bf 102}, 1 (2009).

\bibitem{Gullion1990}
T.~Gullion, D.~B. Baker, M.~S. Conradi, {\it Journal of Magnetic Resonance
  (1969)\/} {\bf 89}, 479 (1990).

\bibitem{deLange2010}
G.~{De Lange}, Z.~Wang, S.~Riste, V.~V. Dobrovitski, R.~Hanson, {\it Science\/}
  {\bf 330}, 60 (2010).

\bibitem{Cruzeiro2017}
E.~Z. Cruzeiro, {\it et~al.\/}, {\it Physical Review B\/} {\bf 95}, 205119
  (2017).

\bibitem{Bottger2006}
T.~B{\"{o}}ttger, C.~W. Thiel, Y.~Sun, R.~L. Cone, {\it Physical Review B\/}
  {\bf 73}, 075101 (2006).

\bibitem{Ranon1968}
U.~Ranon, {\it Physics Letters A\/} {\bf 28}, 228 (1968).

\bibitem{Welinski2017_sc}
S.~Welinski, {\it et~al.\/}, {\it Optical Materials\/} {\bf 63}, 69 (2017).

\end{thebibliography}

\end{document}